\documentclass{shep3}

\usepackage{axodraw,epsfig}
\usepackage{cite}
  \usepackage{graphicx}
  \usepackage{amssymb}

\preprint{
  ZU-TH 07/08\\
  SLAC-PUB-13263
}

\title{Two-Loop Fermionic Corrections
to Heavy-Quark Pair Production:
the Quark-Antiquark Channel}

\author{R.~Bonciani$\rm \, ^{a, \,}$\footnote{Email: {\tt
Roberto.Bonciani@physik.uzh.ch}},  A.~Ferroglia$\rm \, ^{a,
\,}$\footnote{Email: {\tt
Andrea.Ferroglia@physik.uzh.ch}},  T.~Gehrmann$\rm \, ^{a,
\,}$\footnote{Email: {\tt Thomas.Gehrmann@physik.uzh.ch}},
 D.~Ma\^itre$\rm \, ^{b,
\,}$\footnote{Email: {\tt maitreda@slac.stanford.edu}}, and
C.~Studerus$\rm \, ^{a,\,}$\footnote{Email: {\tt cedric@physik.uzh.ch}} \\

{\it $\rm ^a$ Institut f{\"u}r
Theoretische Physik,
Universit{\"a}t Z\"urich,
CH-8057 Zurich, Switzerland}\\

{\it $\rm ^b$ Stanford Linear Accelerator Center,
Stanford University,
Stanford, CA 94390, USA}\\

}

\abstract{We evaluate the fermionic two-loop QCD corrections to  the
heavy-quark pair  production process in the quark-antiquark channel. We obtain
analytic results  which are valid for any  value of the Mandelstam invariants
$s$ and $t$, and of the heavy quark mass $m$.  Our findings
confirm previous results for the analytic evaluation in the small-mass limit
and numerical results for the exact amplitude.
We furthermore provide the expansion
 of the two-loop amplitude at the production
threshold $s \gtrsim 4m^2$.
}


\begin{document}

\newcommand{\be}{\begin{equation}}
\newcommand{\ee}{\end{equation}}
\newcommand{\bfm}[1]{\mbox{\boldmath$#1$}}
\newcommand{\bff}[1]{\mbox{\scriptsize\boldmath${#1}$}}
\newcommand{\al}{\alpha}
\newcommand{\bt}{\beta}
\newcommand{\lm}{\lambda}
\newcommand{\bea}{\begin{eqnarray}}
\newcommand{\eea}{\end{eqnarray}}
\newcommand{\gm}{\gamma}
\newcommand{\Gm}{\Gamma}
\newcommand{\dl}{\delta}
\newcommand{\Dl}{\Delta}
\newcommand{\ep}{\varepsilon}
\newcommand{\vep}{\varepsilon}
\newcommand{\kp}{\kappa}
\newcommand{\Lm}{\Lambda}
\newcommand{\om}{\omega}
\newcommand{\pa}{\partial}
\newcommand{\nn}{\nonumber}
\newcommand{\dd}{\mbox{d}}
\newcommand{\grtsim}{\mbox{\raisebox{-3pt}{$\stackrel{>}{\sim}$}}}
\newcommand{\lessim}{\mbox{\raisebox{-3pt}{$\stackrel{<}{\sim}$}}}
\newcommand{\uk}{\underline{k}}
\newcommand{\gsim}{\;\rlap{\lower 3.5 pt \hbox{$\mathchar \sim$}} \raise 1pt \hbox {$>$}\;}
\newcommand{\lsim}{\;\rlap{\lower 3.5 pt \hbox{$\mathchar \sim$}} \raise 1pt \hbox {$<$}\;}
\newcommand{\Li}{\mbox{Li}}
\newcommand{\bc}{\begin{center}}
\newcommand{\ec}{\end{center}}

\def\lapprox{\lower .7ex\hbox{$\;\stackrel{\textstyle <}{\sim}\;$}}
\def\gapprox{\lower .7ex\hbox{$\;\stackrel{\textstyle >}{\sim}\;$}}

\newcommand{\hypF}{{}_2\mbox{F}_1}


\section{Introduction}
The top quark is the heaviest fermion of the Standard Model. By
studying its properties in detail, it is hoped to elucidate the
origin of particle masses and the mechanism of electroweak
symmetry breaking. Since its discovery at the Fermilab
Tevatron~\cite{toptev} a little more than a decade ago, its mass
has been measured to within a few per cent, while its production
cross section and couplings are only known with larger
uncertainty. With the large number of top quarks expected to be
produced at the LHC, the study of its properties will become
precision physics. To interpret these upcoming precision data,
equally precise theoretical predictions are mandatory. These demand
foremost the calculation of higher order corrections in
perturbative QCD.

At present, the top quark pair production cross section~\cite{Nason:1987xz,Nason:1989zy,Beenakker:1988bq,Beenakker:1990maa,Mangano:1991jk,Korner:2002hy,Bernreuther:2004jv}
 is known to
next-to-leading order (NLO) in the QCD coupling constant,
the same precision is available
for single top production~\cite{Harris:2002md}, $t\bar t$+jet
production~\cite{Dittmaier:2007wz} and
 $t\bar t$+$Z$-boson production~\cite{Lazopoulos:2008de}.
For the pair production cross section,
resummation~\cite{Kidonakis:1997gm,Bonciani:1998vc,Cacciari:2003fi}
 of logarithmically enhanced corrections (next-to-leading
logarithm, NLL) to all orders in the coupling constant improves upon the
fixed-order NLO prediction.
Electroweak one-loop corrections to
$t\bar t$ production are equally available
\cite{Kuhn:2005it,Bernreuther:2006vg}.

Especially for the top quark pair production cross section, which is expected
to be measured to within a few per cent accuracy, it is believed that the
current NLO+NLL prediction is not yet sufficiently accurate. Detailed
recent studies~\cite{Moch:2008qy,Cacciari:2008zb,Kidonakis:2008mu} indicate a
scale uncertainty on these predictions of 7\%, and a parton distribution
uncertainty of 6\%. While the latter may be improved upon by more precise
determinations of the parton distribution functions in view of recent and
upcoming data
from HERA and LHC, the former requires the calculation of
perturbative corrections at next-to-next-to-leading order (NNLO)
in QCD. By approximating these corrections with the fixed-order expansion
of the NLL prediction, one finds~\cite{Moch:2008qy}
a projected NNLO scale uncertainty of
3\%, which is below the parton distribution
uncertainty, and in line with the anticipated
experimental error.

The calculation of the full NNLO corrections to the top quark pair production
cross section requires three types of ingredients:
two-loop matrix elements for $q\bar q \to t\bar t$ and $gg\to t\bar t$,
one-loop matrix elements for hadronic production of
$t\bar t+$(1 parton)  and
tree-level matrix elements for hadronic production of
$t\bar t+$(2 partons). The latter two ingredients were computed
previously in the
context of the NLO corrections to $t\bar t$+jet
production~\cite{Dittmaier:2007wz}. They
contribute to the  $t\bar t$ production cross section through configurations
where up to two final state partons can be unresolved (collinear or soft),
and their implementation thus may require further developments of
subtraction techniques at NNLO.

Both two-loop matrix elements were computed
analytically
in the small-mass expansion
limit
$s, |t|, |u| \gg m^2$ in~\cite{Czakon:2007ej,Czakon:2007wk},
starting from the previously known massless two-loop
  matrix elements for $q\bar q\to q'\bar q'$~\cite{Anastasiou:2000kg}
 and $gg\to q\bar q$~\cite{Anastasiou:2001sv}. An exact numerical
representation of
the two-loop matrix element  $q\bar q \to t\bar t$ has been obtained very
recently~\cite{Czakon:2008zk}. It is the aim of the present paper to
compute all two-loop contributions to
 $q\bar q \to t\bar t$ arising from closed fermion loops in a compact analytic
form, which provide a first independent validation of the recent results
of~\cite{Czakon:2007ej,Czakon:2008zk},
 allow for a fast numerical evaluation, and
permit the analytical study of the behavior of the top quark
production cross section at threshold.

This paper is structured as follows. In Section~\ref{nota}, we
define our notation and kinematical conventions.
Sections~\ref{sec:calc} and~\ref{sec:renorm}
describe the details of the calculation of the two-loop integrals  and
of the renormalization of the amplitudes. The results are presented and
discussed in Section~\ref{sec:results}. We enclose two appendices describing
the special functions used in our calculation and documenting the newly
computed master integrals.


\begin{figure}
\vspace*{.5cm}
\[ \hspace*{-3mm} \vcenter{ \hbox{
  \begin{picture}(0,0)(0,0)
\SetScale{1}
  \SetWidth{.5}
\ArrowLine(-55,30)(-30,0)
\ArrowLine(-30,0)(-55,-30)
\Gluon(-30,0)(30,0){3}{10}
\LongArrow(-62,24)(-50,10)
\LongArrow(-62,-24)(-50,-10)
\LongArrow(50,10)(62,24)
\LongArrow(50,-10)(62,-24)
  \SetWidth{1.6}
\ArrowLine(55,30)(30,0)
\ArrowLine(30,0)(55,-30)

\Text(-67,27)[cb]{$p_1$}
\Text(-67,-33)[cb]{$p_2$}
\Text(69,-33)[cb]{$p_3$}
\Text(69,27)[cb]{$p_4$}
\end{picture}}
}
\]
 \vspace*{.6cm} \caption{\it Tree-level  amplitude. Massive
quarks are indicated by a thick line.} \label{figTree}
\end{figure}
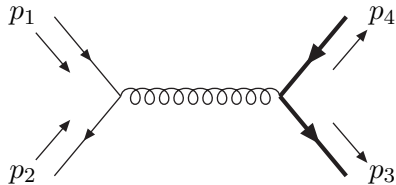


\section{Notation and Conventions \label{nota}}

We consider the scattering process \be q(p_1) + \overline{q}(p_2)
\longrightarrow  t(p_3) + \overline{t}(p_4) \, , \ee
in Euclidean kinematics, where $p_i^2 = 0$ for $i=1,2$  and
$p_j^2 = -m^2$ for  $i=3,4$. The Mandelstam variables are defined as
follows
\be
s = -\left(p_1 + p_2 \right)^2 \, , \quad
t = -\left(p_1 - p_3 \right)^2 \, , \quad
u = -\left(p_1 - p_4 \right)^2 \, .
\ee
Conservation of momentum implies that $s +t +u = 2 m^2$.

The squared tree-level matrix element (averaged over the spin and color of the
incoming quarks and summed over the spin of the outgoing ones), calculated in
$d = 4 -2 \varepsilon$ dimensions, can be expanded in powers of the strong
coupling constant $\alpha_S$ as follows:
\be \label{M2}
|\mathcal{M}|^2(s,t,m,\varepsilon) = \frac{4 \pi^2 \alpha_S^2}{N_c^2}
\left[{\mathcal A}_0 +
\left(\frac{\alpha_s}{ \pi} \right) {\mathcal A}_1 +
\left(\frac{\alpha_s}{ \pi} \right)^2 {\mathcal A}_2 +
{\mathcal O}\left( \alpha_s^3\right)\right] \, .
\ee
The tree-level amplitude involves a single diagram (Fig.~\ref{figTree}) and its contribution to Eq.~(\ref{M2}) is given by
\be
{\mathcal A}_0=  4  N_c \, C_F
\left[ \frac{(t-m^2)^2+(u-m^2)^2}{s^2} + \frac{2 m^2}{s} - \varepsilon\right] \, ,
\label{treeM}
\ee
where  $N_c$ is the number of colors and  $C_F = (N_c^2-1)/2N_c$.
As it is well known, the term proportional to the dimensional regulator
$\varepsilon$ in Eq.~(\ref{treeM}) is mass independent.

The NLO term ${\mathcal A}_1$  in Eq.~(\ref{M2}) arises from the interference
of one-loop diagrams with the tree-level amplitude
\cite{Nason:1987xz,Nason:1989zy,Beenakker:1988bq,Beenakker:1990maa,Mangano:1991jk,Korner:2002hy,Bernreuther:2004jv}.
The NNLO term ${\mathcal A}_2$ consists of two parts, the interference of
two-loop diagrams with the Born amplitude and the interference of one-loop
diagrams among themselves:
\begin{displaymath}
{\mathcal A}_2 = {\mathcal A}_2^{(2\times 0)} + {\mathcal A}_2^{(1\times 1)}\;.
\end{displaymath}
The latter term ${\mathcal A}_2^{(1\times 1)}$ was
studied  extensively  in
\cite{Korner:2005rg}. ${\mathcal A}_2^{(2\times 0)}$,  originating
from the two-loop  diagrams, can be decomposed according to color and
flavor structures as follows:
\bea  {\mathcal A}_2^{(2\times 0)}  &=&  N_c C_F \Biggl[ N_c^2 A
+B +\frac{C}{N_c^2}  + N_l \left( N_c D_l + \frac{E_l}{N_c}
\right)
+ N_h \left( N_c D_l + \frac{E_l}{N_c} \right)  \nn \\
& & \hspace*{10mm} + N_l^2 F_l + N_l N_h F_{lh} + N_h^2 F_h\Biggr] \, ,
\label{colstruc}
\eea
where $N_l$ and $N_h$ are the number of light- and heavy-quark flavors,
respectively. The coefficients $A,B,\ldots,F_h$ in Eq.~(\ref{colstruc}) are
functions of $s$, $t$, and $m$, as well as of the dimensional regulator
$\varepsilon$.
Recently, these quantities were calculated in  \cite{Czakon:2007ej} in the
 approximation $s,|t|,|u| \gg m^2$. For a fully
differential description of top quark pair production at NNLO, the
complete mass dependence of  ${\mathcal A}_2^{(2\times 0)}$ is required.
An exact numerical expression for it has been obtained very recently in
\cite{Czakon:2008zk}. In this work, we provide independent
confirmations of the recent results of  \cite{Czakon:2007ej,Czakon:2008zk}
by deriving exact analytic  expressions for all the terms in
Eq.~(\ref{colstruc}) arising from  two-loop diagrams involving at least a
fermion loop (i.e. the  coefficients $D_i,E_i,F_j$ with $i=l,h$ and $j=l,h,lh$).

\section{Calculation\label{sec:calc}}


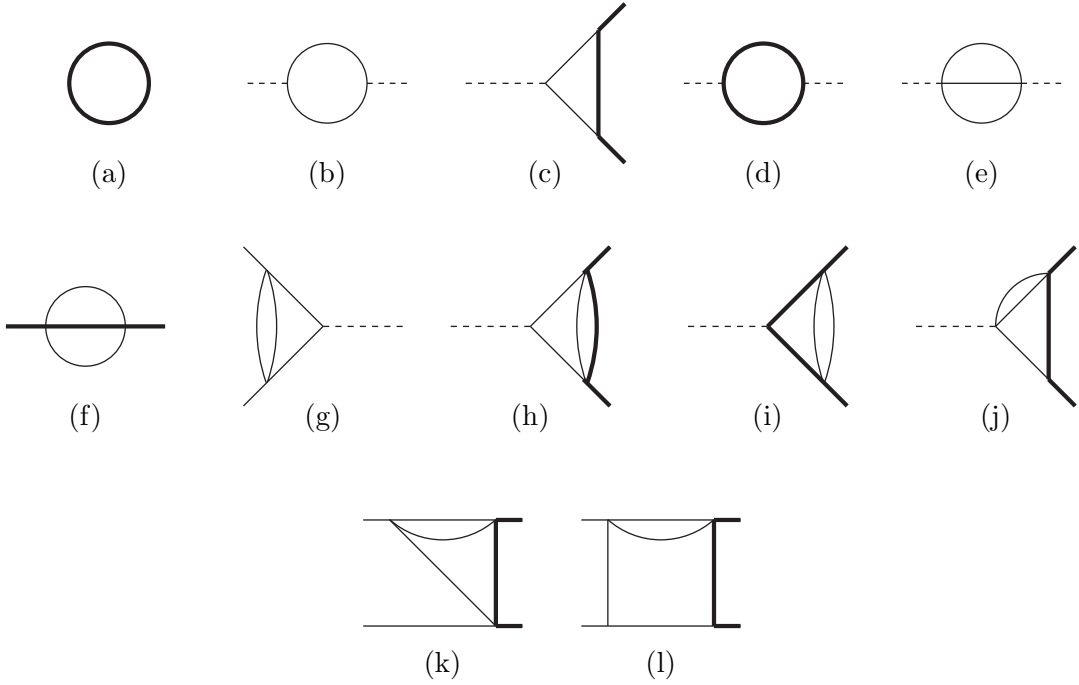
\begin{figure}
\vspace*{.5cm}
\[ \hspace*{-3mm}
\vcenter{
\hbox{ \begin{picture}(0,0)(0,0)
\SetScale{1}
  \SetWidth{.5}
\Text(0,-40)[cb]{(a)}
  \SetWidth{1.6}
\CArc(0,0)(15,0,360)
\end{picture}}
}
\hspace{2.9cm}
  \vcenter{
\hbox{\begin{picture}(0,0)(0,0)
\SetScale{1}
  \SetWidth{.5}
\DashLine(-30,0)(-15,0){2.5} \DashLine(15,0)(30,0){2.5}
\CArc(0,0)(15,0,360)
\Text(0,-40)[cb]{(b)}
  \SetWidth{1.6}
\end{picture}}
}
\hspace{2.9cm}
  \vcenter{
\hbox{\begin{picture}(0,0)(0,0)
\SetScale{1}
  \SetWidth{.5}
  \DashLine(-30,0)(0,0){2.5}
  \Line(0,0)(30,30)
  \Line(0,0)(30,-30)
\Text(0,-40)[cb]{(c)}
  \SetWidth{1.6}
  \Line(20,20)(30,30)
  \Line(20,20)(20,-20)
  \Line(20,-20)(30,-30)
\end{picture}}
}
%
%
\hspace{2.9cm}
  \vcenter{
\hbox{\begin{picture}(0,0)(0,0)
\SetScale{1}
  \SetWidth{.5}
\DashLine(-30,0)(-15,0){2.5} \DashLine(15,0)(30,0){2.5}
\Text(0,-40)[cb]{(d)}
  \SetWidth{1.6}
\CArc(0,0)(15,0,360)
\end{picture}}
}
\hspace{2.9cm}
\vcenter{ \hbox{\begin{picture}(0,0)(0,0) \SetScale{1}
  \SetWidth{.5}
\Line(-15,0)(15,0) \DashLine(-30,0)(-15,0){2.5}
\DashLine(15,0)(30,0){2.5} \CArc(0,0)(15,0,360)
\Text(0,-40)[cb]{(e)}
  \SetWidth{1.6}
\end{picture}}}
\]
\vspace*{1.8cm}
\[ \hspace*{-3mm}
  \vcenter{
\hbox{\begin{picture}(0,0)(0,0)
\SetScale{1}
  \SetWidth{.5}
\CArc(0,0)(15,0,360)
\Text(0,-40)[cb]{(f)}
  \SetWidth{1.6}
\Line(-30,0)(30,0)
\end{picture} }
}
\hspace{2.9cm}
  \vcenter{
\hbox{ \begin{picture}(0,0)(0,0)
\SetScale{1}
  \SetWidth{.5}
\DashLine(0,0)(30,0){2.5}
\Line(0,0)(-30,30)
\Line(0,0)(-30,-30)
\CArc(40,0)(65,161,199)
\CArc(-82.5,0)(65,-19,19)
\Text(0,-40)[cb]{(g)}
  \SetWidth{1.6}
\end{picture} }
}
\hspace{2.5cm}
  \vcenter{
\hbox{ \begin{picture}(0,0)(0,0)
\SetScale{1}
  \SetWidth{.5}
\DashLine(0,0)(-30,0){2.5}
\Line(0,0)(30,30)
\Line(0,0)(30,-30)
\CArc(82.5,0)(65,161,199)
\Text(0,-40)[cb]{(h)}
  \SetWidth{1.6}
\Line(20,20)(30,30)
\Line(20,-20)(30,-30)
\CArc(-40,0)(65,-19,19)
\end{picture} }
}
\hspace{2.9cm}
  \vcenter{
\hbox{ \begin{picture}(0,0)(0,0)
\SetScale{1}
  \SetWidth{.5}
\DashLine(0,0)(-30,0){2.5}
\CArc(82.5,0)(65,161,199)
\CArc(-40,0)(65,-19,19)
\Text(0,-40)[cb]{(i)}
  \SetWidth{1.6}
\Line(0,0)(30,30)
\Line(0,0)(30,-30)
\end{picture} }}
\hspace{2.9cm}
\vcenter{\hbox{\begin{picture}(0,0)(0,0) \SetScale{1}
  \SetWidth{.5}
  \DashLine(-30,0)(0,0){2.5}
  \Line(0,0)(30,30)
  \Line(0,0)(30,-30)
  \CArc(20,0)(20,90,180)
\Text(0,-40)[cb]{(j)}
  \SetWidth{1.6}
  \Line(20,20)(30,30)
  \Line(20,20)(20,-20)
  \Line(20,-20)(30,-30)
\end{picture}}}
\]
\vspace*{2.0cm}
\[
%
\vcenter{\hbox{\begin{picture}(0,0)(0,0)
\SetScale{1}
  \SetWidth{.5}
  \Line(-30,20)(30,20)
  \Line(-30,-20)(30,-20)
  \Line(-20,20)(20,-20)
\CArc(0,42.5)(30,228,312)
\Text(0,-40)[cb]{(k)}
  \SetWidth{1.6}
  \Line(20,20)(30,20)
  \Line(20,-20)(30,-20)
  \Line(20,20)(20,-20)
\end{picture}}
  }
\hspace{2.9cm}
\vcenter{\hbox{\begin{picture}(0,0)(0,0)
\SetScale{1}
  \SetWidth{.5}
  \Line(-30,20)(30,20)
  \Line(-30,-20)(30,-20)
  \Line(-20,20)(-20,-20)
\CArc(0,42.5)(30,228,312)
\Text(0,-40)[cb]{(l)}
  \SetWidth{1.6}
  \Line(20,20)(30,20)
  \Line(20,-20)(30,-20)
  \Line(20,20)(20,-20)
\end{picture}}
  }
\]
\vspace*{.8cm}
\caption{\it Non-reducible topologies for the light quark corrections.}
\label{nlMIs}
\end{figure}



\begin{figure}
\vspace*{.5cm}
\[ \hspace*{-3mm}
\vcenter{
\hbox{\begin{picture}(0,0)(0,0)
\SetScale{1}
  \SetWidth{.5}
  \Line(-25,20)(0,20)
  \Line(-25,-20)(0,-20)
  \CArc(15,0)(25,127,233)
\Text(0,-40)[cb]{(a)}
  \SetWidth{1.6}
  \Line(25,20)(0,20)
  \Line(25,-20)(0,-20)
  \CArc(-15,0)(25,-53,53)

\end{picture} }
}
\hspace{2.9cm}
  \vcenter{
\hbox{\begin{picture}(0,0)(0,0)
\SetScale{1}
  \SetWidth{.5}
  \Line(-30,20)(30,20)
  \Line(-30,-20)(30,-20)
  \Line(-20,20)(-20,-20)
\Text(0,-40)[cb]{(b)}
  \SetWidth{1.6}
  \Line(20,20)(30,20)
  \Line(20,20)(20,-20)
  \Line(20,-20)(30,-20)
\end{picture}}
}
\hspace{2.9cm}
  \vcenter{
\hbox{\begin{picture}(0,0)(0,0)
\SetScale{1}
  \SetWidth{.5}
\Line(-15,0)(15,0)
\DashLine(-30,0)(-15,0){2.5}
\DashLine(30,0)(15,0){2.5}
\Text(0,-40)[cb]{(c)}
  \SetWidth{1.6}
\CArc(0,0)(15,0,360)
\end{picture}}
}
\hspace{2.9cm}
  \vcenter{
\hbox{\begin{picture}(0,0)(0,0)
\SetScale{1}
  \SetWidth{.5}
\DashLine(-30,0)(-15,0){2.5}
\DashLine(15,0)(30,0){2.5}
\CArc(0,0)(15,0,360)
\Text(0,-40)[cb]{(d)}
  \SetWidth{1.6}
\CArc(0,0)(15,90,270)
\Line(0,15)(0,-15)
\end{picture}}
}
\hspace{2.9cm}
  \vcenter{
\hbox{\begin{picture}(0,0)(0,0)
\SetScale{1}
  \SetWidth{.5}
\DashLine(0,0)(30,0){2.5}
\Line(0,0)(-30,30)
\Line(0,0)(-30,-30)
\Text(0,-40)[cb]{(e)}
  \SetWidth{1.6}
\CArc(40,0)(65,161,199)
\CArc(-82.5,0)(65,-19,19)
\end{picture}}
}
\]\vspace*{2.0cm}
\[
\vcenter{\hbox{\begin{picture}(0,0)(0,0)
\SetScale{1}
  \SetWidth{.5}
\DashLine(0,0)(30,0){2.5}
\Line(0,0)(-30,30)
\Line(0,0)(-30,-30)
\Text(0,-40)[cb]{(f)}
\CArc(40,0)(65,161,199)
  \SetWidth{1.6}
\Line(0,0)(-20,20)
\Line(0,0)(-20,-20)
\CArc(-82.5,0)(65,-19,19)
\end{picture}}}
%
\hspace{2.9cm}
\vcenter{\hbox{\begin{picture}(0,0)(0,0)
\SetScale{1}
  \SetWidth{.5}
\Text(0,-40)[cb]{(g)}
  \SetWidth{1.6}
\CArc(0,0)(15,0,360)
\Line(-30,0)(30,0)
\end{picture}}
  }
\hspace{2.9cm}
\vcenter{\hbox{\begin{picture}(0,0)(0,0)
\SetScale{1}
  \SetWidth{.5}
\DashLine(0,0)(-30,0){2.5}
\Text(0,-40)[cb]{(h)}
  \SetWidth{1.6}
\Line(0,0)(30,30)
\Line(0,0)(30,-30)
\CArc(82.5,0)(65,161,199)
\CArc(-40,0)(65,-19,19)
\end{picture}}
  }
\hspace{2.9cm}
\vcenter{\hbox{\begin{picture}(0,0)(0,0)
\SetScale{1}
  \SetWidth{.5}
  \DashLine(-30,0)(0,0){2.5}
  \Line(0,0)(30,-30)
\Text(0,-40)[cb]{(i)}
  \SetWidth{1.6}
  \CArc(20,0)(20,90,180)
  \Line(0,0)(30,30)
  \Line(20,20)(30,30)
  \Line(20,20)(20,-20)
  \Line(20,-20)(30,-30)
\end{picture}}
  }
\hspace{2.9cm}
\vcenter{\hbox{\begin{picture}(0,0)(0,0)
\SetScale{1}
  \SetWidth{.5}
  \DashLine(-30,0)(0,0){2.5}
  \Line(0,0)(30,-30)
  \Line(0,0)(30,30)
\Text(0,-40)[cb]{(j)}
  \SetWidth{1.6}
  \Line(10,10)(10,-10)
  \Line(0,0)(10,10)
  \Line(0,0)(10,-10)
  \Line(20,20)(30,30)
  \Line(20,20)(20,-20)
  \Line(20,-20)(30,-30)
\end{picture}}
  }
\]
\vspace*{2.0cm}
\[
%
\vcenter{\hbox{\begin{picture}(0,0)(0,0)
\SetScale{1}
  \SetWidth{.5}
  \Line(-30,20)(30,20)
  \Line(-30,-20)(30,-20)
  \Line(-20,20)(20,-20)
\Text(0,-40)[cb]{(k)}
  \SetWidth{1.6}
  \Line(20,20)(30,20)
  \Line(20,-20)(30,-20)
  \Line(20,20)(20,-20)
  \Line(-20,20)(20,20)
\CArc(0,42.5)(30,228,312)
\end{picture}}
  }
\hspace{2.9cm}
\vcenter{\hbox{\begin{picture}(0,0)(0,0)
\SetScale{1}
  \SetWidth{.5}
  \Line(-30,20)(30,20)
  \Line(-30,-20)(30,-20)
  \Line(-20,20)(-20,-20)
\Text(0,-40)[cb]{(l)}
  \SetWidth{1.6}
  \Line(20,20)(30,20)
  \Line(-20,20)(20,20)
  \Line(20,-20)(30,-20)
  \Line(20,20)(20,-20)
\CArc(0,42.5)(30,228,312)
\end{picture}}
  }
\]
\vspace*{.8cm}
\caption{\it Non-reducible topologies for the heavy quark corrections.}
\label{nhMIs}
\end{figure}
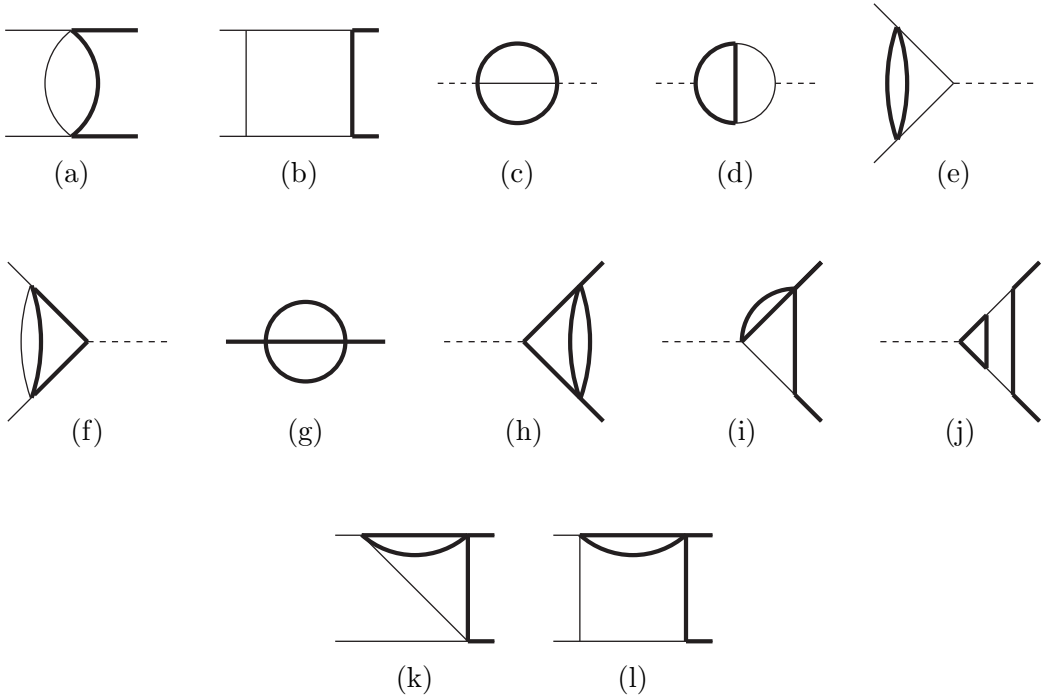

The calculation starts from the two-loop Feynman diagrams
for $q\bar q\to t\bar t$, which are generated using QGRAF~\cite{qgraf},
interfered with the tree-level amplitude, and simplified using
FORM~\cite{FORM}.
Out of the 218 two-loop diagrams contributing to the amplitude, 28 are
proportional to $N_l$, 29 are proportional to $N_h$, 2 are proportional to $N_l
N_h$, while just one contributes to the $N_l^2$ and $N_h^2$ parts. Most
importantly,  there is only one two-loop box topology contributing to the
$N_l$ part of the squared amplitude, and a single other two-loop box topology
proportional to $N_h$. These two box topologies are very similar to
the ones encountered in the evaluation of the two-loop QED corrections to Bhabha
scattering  \cite{electronloop,hfbha}, and can be evaluated with the same
techniques.

All two-loop integrals appearing in these amplitudes
are reduced to a set of master integrals (MIs) by
means of the standard method based on the Laporta algorithm
\cite{Laportaalgorithm}. Only part of these MIs were available
in the
literature~\cite{Argeri:2002wz,Bonciani:2003te,Aglietti:2003yc,DK,Bonciani:2003hc,Aglietti:2004tq,CGR}
from previous two-loop calculations of the heavy quark form
factors~\cite{Bernreuther:2004ih}
and amplitudes for Bhabha scattering~\cite{electronloop,hfbha,bhabha}.
 For the remaining integrals, we employed the  differential equation method
\cite{DiffEq}.

The reduction to MIs was carried out with two independent implementations
of  the Laporta algorithm, and large parts of it were cross checked with the
{\tt Maple} package {\tt A.I.R.} \cite{Anastasiou:2004vj}.  The 12 irreducible
topologies encountered in the calculation of the diagrams with  a light quark
loop are shown in Fig.~\ref{nlMIs}. The diagrams proportional to $N_h$
also contain 12 irreducible topologies, which can be
found in Fig.~\ref{nhMIs}.  In both figures, thick internal  lines indicate
massive propagators, while thin lines indicate massless ones. An external
dashed  leg carries a squared momentum $s$; other external  lines indicate
particles on their mass-shell, where $p_i^2 = 0$ for thin lines and
$p_i^2 = -m^2$ for thick lines.

The analytic expressions of the one-loop MIs in Fig.~\ref{nlMIs} and
Fig.~\ref{nhMIs} are well known. The large majority of the two-loop MIs is also
known in the literature: explicit analytic expressions for all the two-loop MIs
with the exception of the ones belonging to topologies Fig.~\ref{nlMIs}-(k),
Fig.~\ref{nlMIs}-(l),
Fig.~\ref{nhMIs}-(k), and Fig.~\ref{nhMIs}-(l) can be found in
\cite{Argeri:2002wz,Bonciani:2003te,Aglietti:2003yc,Bonciani:2003hc}.

The MIs associated with topologies Fig.~\ref{nlMIs}-(k), Fig.~\ref{nlMIs}-(l),
Fig.~\ref{nhMIs}-(k), and
Fig.~\ref{nhMIs}-(l) that were not available in the literature
 are collected in
Appendix~\ref{AppMIs}. In calculating the MIs by means of the differential
equation method, it is crucial to fix the undetermined integration constant(s)
appearing  in the solution of the differential equations. While there is no
general method available to fix such initial condition, it is usually
sufficient to know the behavior of the MI in some particular kinematic
point; for example, knowing that the integral is regular for a certain
value of $s$, one can impose the regularity of the solution of the differential
equation in that point. This can be sufficient to determine the integration
constant.
In our calculation, the initial conditions for the single master
integral belonging to topology Fig.~\ref{nlMIs}-(k) and the two
MIs belonging to topology Fig.~\ref{nhMIs}-(l) were determined by
imposing the regularity  of the solution of the differential
equation in $t=0$. However, this is not always sufficient. For
topology Fig.~\ref{nhMIs}-(k), which has two MIs, imposing the
regularity of both MIs in $t=0$ allowed to fix only one of the two
initial conditions required. In order to fix the second
integration constant, we had to  use another piece of information,
namely that the scalar integral  with all the denominators raised
to power one diverges at most logarithmically  at the threshold
$t=m^2$. The final result for these two MIs was then checked by
calculating their $t \to 0$ limit with the Mellin Barnes
technique, using the {\tt Mathematica} packages {\tt Ambre}
\cite{Gluza:2007rt} and {\tt MB} \cite{Czakon:2005rk}. For what
concerns the MI of the box topology in Fig.~\ref{nlMIs}-(l), the
initial conditions were defined calculating the integral in $t=0$
with Mellin Barnes techniques.

All the MIs were calculated in the non-physical region $s<0$, where they are
real and can be conveniently written as functions of the dimensionless
variables
\be
x = \frac{\sqrt{-s+4 m^2} - \sqrt{-s}}{\sqrt{-s+4 m^2} + \sqrt{-s}} \, ,
\qquad y = \frac{-t}{m^2} \, , \qquad z = \frac{-u}{m^2} \, .
\ee
The MIs of the topology Fig.~\ref{nhMIs}-(e)
are an exception. In this case it is
convenient to employ the variable
\be \label{xp}
x_p = \frac{\sqrt{-s} - \sqrt{-s-4 m^2}}{\sqrt{-s} + \sqrt{-s-4 m^2}} \, .
\ee
The transcendental functions appearing in the MIs are one- and two-dimensional
harmonic polylogarithms (HPLs) \cite{HPLs,2dHPLs}. In the result one finds
one-dimensional  HPLs of maximum weight four and two-dimensional HPLs of
maximum  weight three. Both  sets of functions can be rewritten in  terms of
conventional Nielsen's  polylogarithms. In Appendix~\ref{AppGPLs}, we briefly
review the definition of  the HPLs employed and we collect the expression of
some of them in  terms of Nielsen's polylogarithms.

Following the procedure outlined in the present section, it is possible to
obtain the expression of the bare squared matrix elements involving diagrams
proportional to $N_l$ and/or $N_h$. After this goal is achieved, it is then
necessary to renormalize the ultraviolet divergencies.
In the next section, we briefly discuss the renormalization procedure and we
explicitly list the needed renormalization constants.


\section{Renormalization\label{sec:renorm}}

The renormalized QCD matrix element is obtained from the bare one
by expanding the following expression :
\be
{\mathcal A}_{{\small \mbox{ren}}} = \prod_{n} Z^{1/2}_{{\tiny \mbox{WF}},n} \,
{\mathcal A}_{{\small \mbox{bare}}}
\left( \alpha_{S, {\small \mbox{bare}}} \to Z_{\alpha_S} \alpha_S \, ,
m_{{\small \mbox{bare}}} \to Z_m m \right) \, ,
\label{renM}
\ee
where $Z_{{\tiny \mbox{WF}},n}$ is the external leg wave function renormalization
factor, $\alpha_S$ is the renormalized coupling constant and $m$ is the renormalized
heavy quark mass. (In the rest of the section we suppress the subscript ``$S$'' in
$\alpha_S$).

We postpone the discussion of mass renormalization to the end of the section and we
start by  considering the coupling constant and wave function renormalization.

We introduce the following quantities:
\be
a_0 = \frac{\alpha_{\small \mbox{bare}}}{\pi}  \, ,
\qquad \mbox{and} \qquad
a = \frac{\alpha}{\pi}  \, .
\ee
By expanding the amplitude and the wave function renormalization factor
in $a_0$ we find:
\bea
{\mathcal A}_{{\small \mbox{ren}}}(\alpha_{\small \mbox{bare}}) &=&
    a_0 {\mathcal A}_0
  + a_0^2 {\mathcal A}_1
  + a_0^3 {\mathcal A}_2
  + {\mathcal O}(a_0^4) \, , \nn \\
Z_{{\tiny \mbox{WF}},n} & = & 1
  + a_0 \delta Z^{(1)}_{{\tiny \mbox{WF}},n}
  + a_0^2 \delta Z^{(2)}_{{\tiny \mbox{WF}},n}
  + {\mathcal O}(a_0^3) \, .\label{exp1}
\eea
The relation between $a_0$ and $a$ is given by:
\be
a_0  =  a
  + a^2 \delta Z^{(1)}_{\alpha}
  + a^3\delta Z^{(2)}_{\alpha}
  + {\mathcal O}(a^4) \, . \label{exp2}
\ee
By employing  Eqs.~(\ref{exp1},\ref{exp2}) in Eq.~(\ref{renM}) we find
\bea
{\mathcal A}_{{\small \mbox{ren}}} &=& a {\mathcal A}_0
+ a^2 {\mathcal A}^{(1)} _{{\small \mbox{ren}}} +
a^3 {\mathcal A}^{(2)} _{{\small \mbox{ren}}} + \mathcal{O}(a^4)\, , \nn \\
{\mathcal A}^{(1)} _{{\small \mbox{ren}}} &=& {\mathcal A}_1 +
\left(\sum_n \frac{1}{2} \delta Z^{(1)}_{{\tiny \mbox{WF}},n} + \delta
Z^{(1)}_{\alpha}\right)  {\mathcal A}_0 \, , \nn \\
{\mathcal A}^{(2)} _{{\small \mbox{ren}}}&=&   {\mathcal A}_2 +
 \left(\sum_n \frac{1}{2} \delta Z^{(1)}_{{\tiny \mbox{WF}},n}  + 2 \delta
Z^{(1)}_{\alpha} \right) {\mathcal A}_1+
\left(-\sum_n \frac{1}{8} \left(\delta Z^{(1)}_{{\tiny \mbox{WF}},n}\right)^2  \right.   \nn \\
& & + \left.\sum_n \frac{1}{2} \delta Z^{(2)}_{{\tiny \mbox{WF}},n} +
  \delta Z^{(1)}_{\alpha}\sum_n \delta Z^{(1)}_{{\tiny \mbox{WF}},n}
 +\delta Z^{(2)}_{\alpha}
\right) {\mathcal A}_0\, .\label{exprenM}
\eea
In the equations above, ${\mathcal A}_i$ represents the amplitude at $i$ loops
stripped of the factor $a$. In the case of the process $ q \overline{q} \to t
\overline{t}$, the wave function renormalization factors of  massless quarks vanish at
one loop, while the ones of the massive quarks in the on-shell renormalization scheme
are given by
\be
\delta Z^{(1)}_{{\tiny \mbox{WF}},M} = C(\vep) \,
             \left(\frac{\mu^2}{m^2} \right)^{\vep} C_F
         \left( -\frac{3}{4 \ep} - \frac{1}{1-2 \ep}  \right) \, ,
\ee
where the subscript $M$ indicates massive quarks and where $C(\vep) = (4 
\pi)^{\vep}
\Gamma(1+\vep)$.
The one-loop renormalization constant for $\alpha$  in the $\overline{MS}$ scheme is
given by
\be
\delta Z^{(1)}_{\alpha} = C(\vep) \,
\frac{e^{-\gamma \vep}}{\Gamma(1+\vep)} \left( -\frac{\beta_0}{2 \vep}\right) \, ,
\ee
where $\beta_0 = 11/6 C_A  -1/3 (N_l + N_h)$ and $\gamma$ is the Euler-Mascheroni
constant $\gamma \approx  0.577216$.

Therefore, the overall one-loop counter term is given by
\be
\delta Z^{(1)}_{{\tiny \mbox{WF}},M} + \delta Z^{(1)}_{\alpha} =
-\frac{C(\vep)}{4 \ep} \left[ 2 \beta_0+ 3 + 4 \vep + \ln{\left(\frac{\mu^2}{m^2}\right)} \right] + \mathcal{O} \left( \vep^2 \right) \,.
\ee
%

\begin{figure}
\vspace*{.5cm}
\[ \hspace*{-3mm}
\vcenter{
\hbox{ \begin{picture}(0,0)(0,0)
\SetScale{1}
  \SetWidth{.5}
\ArrowLine(-55,30)(-30,0) \ArrowLine(-30,0)(-55,-30)
\Gluon(-30,0)(-15,0){3}{2.5} \Gluon(15,0)(30,0){3}{2.5}
\ArrowArc(0,0)(15,0,180) \ArrowArc(0,0)(15,180,360)
\BText(0,-45){1l-1}
  \SetWidth{1.6}
\ArrowLine(30,0)(55,30)
\ArrowLine(55,-30)(30,0)
\end{picture}}
}
\hspace{4.6cm}
  \vcenter{
\hbox{\begin{picture}(0,0)(0,0)
\SetScale{1}
  \SetWidth{.5}
\ArrowLine(-55,30)(-30,0) \ArrowLine(-30,0)(-55,-30)
\Gluon(-30,0)(-15,0){3}{2.5} \Gluon(15,0)(30,0){3}{2.5}
\BText(0,-45){1l-2}
  \SetWidth{1.6}
\ArrowLine(30,0)(55,30)
\ArrowLine(55,-30)(30,0)
\ArrowArc(0,0)(15,0,180)
\ArrowArc(0,0)(15,180,360)
\end{picture}}
}
\hspace{4.6cm}
  \vcenter{
\hbox{\begin{picture}(0,0)(0,0)
\SetScale{1}
  \SetWidth{.5}
\ArrowLine(-55,30)(-30,0) \ArrowLine(-30,0)(-55,-30)
\Gluon(-30,0)(-15,0){3}{2.5} \Gluon(15,0)(30,0){3}{2.5}
\GlueArc(0,0)(15,0,180){3}{8} \GlueArc(0,0)(15,180,360){3}{8}
\BText(0,-45){1l-3}
  \SetWidth{1.6}
\ArrowLine(30,0)(55,30)
\ArrowLine(55,-30)(30,0)
\end{picture}}
}
\]
\vspace*{2.3cm}
\[ \hspace*{-3mm}
\vcenter{
\hbox{ \begin{picture}(0,0)(0,0)
\SetScale{1}
  \SetWidth{.5}
\ArrowLine(-55,30)(-30,0) \ArrowLine(-30,0)(-55,-30)
\Gluon(-30,0)(-15,0){3}{2.5} \Gluon(15,0)(30,0){3}{2.5}
\DashArrowArcn(0,0)(15,0,180){3}
\DashArrowArcn(0,0)(15,180,360){3}
\BText(0,-45){1l-4}
  \SetWidth{1.6}
\ArrowLine(30,0)(55,30)
\ArrowLine(55,-30)(30,0)
\end{picture}}
}
\hspace{4.6cm}
  \vcenter{
\hbox{\begin{picture}(0,0)(0,0)
\SetScale{1}
  \SetWidth{.5}
\ArrowLine(-55,30)(-30,0) \ArrowLine(-30,0)(-55,-30)
\Gluon(-30,0)(0,0){3}{4.5} \Gluon(0,0)(40,30){3}{7.5}
\Gluon(0,0)(40,-30){3}{7.5} \BText(0,-45){1l-5}
  \SetWidth{1.6}
\ArrowLine(40,30)(55,30)
\ArrowLine(55,-30)(40,-30)
\ArrowLine(40,-30)(40,30)
\end{picture}}
}
\hspace{4.6cm}
  \vcenter{
\hbox{\begin{picture}(0,0)(0,0)
\SetScale{1}
  \SetWidth{.5}
\ArrowLine(-55,30)(-30,0) \ArrowLine(-30,0)(-55,-30)
\Gluon(-30,0)(0,0){3}{4.5} \Gluon(40,-30)(40,30){3}{10.5}
\BText(0,-45){1l-6}
  \SetWidth{1.6}
\ArrowLine(40,30)(55,30)
\ArrowLine(55,-30)(40,-30)
\ArrowLine(0,0)(40,30)
\ArrowLine(40,-30)(0,0)
\end{picture}}
}
\]
\vspace*{2.3cm}
\[
\vcenter{\hbox{\begin{picture}(0,0)(0,0)
\SetScale{1}
  \SetWidth{.5}
\Gluon(30,0)(0,0){3}{4.5} \ArrowLine(-40,-30)(-40,30)
\ArrowLine(-40,30)(-55,30) \ArrowLine(-40,-30)(-55,-30)
\Gluon(-40,30)(0,0){3}{7.5} \Gluon(0,0)(-40,-30){3}{7.5}
\BText(0,-45){1l-7}
  \SetWidth{1.6}
\ArrowLine(30,0)(55,30)
\ArrowLine(55,-30)(30,0)
\end{picture}}}
\hspace{4.6cm}
\vcenter{\hbox{\begin{picture}(0,0)(0,0)
\SetScale{1}
  \SetWidth{.5}
\Gluon(30,0)(0,0){3}{4.5} \Gluon(-40,-30)(-40,30){3}{10.5}
\ArrowLine(-55,30)(-40,30) \ArrowLine(-40,-30)(-55,-30)
\ArrowLine(-40,30)(0,0) \ArrowLine(0,0)(-40,-30)
\BText(0,-45){1l-8}
  \SetWidth{1.6}
\ArrowLine(30,0)(55,30)
\ArrowLine(55,-30)(30,0)
\end{picture}}
  }
\]
\vspace*{2.3cm}
\[
\vcenter{\hbox{\begin{picture}(0,0)(0,0)
\SetScale{1}
  \SetWidth{.5}
\ArrowLine(-40,-30)(-40,30) \ArrowLine(-40,30)(-55,30)
\ArrowLine(-40,-30)(-55,-30) \Gluon(-40,30)(40,30){3}{12.5}
\Gluon(40,-30)(-40,-30){3}{12.5} \BText(0,-50){1l-9}
  \SetWidth{1.6}
\ArrowLine(40,-30)(40,30)
\ArrowLine(40,30)(55,30)
\ArrowLine(55,-30)(40,-30)
\end{picture}}}
\hspace{4.6cm}
\vcenter{\hbox{\begin{picture}(0,0)(0,0)
\SetScale{1}
  \SetWidth{.5}
\ArrowLine(-40,-30)(-40,30)
\ArrowLine(-40,30)(-55,30)
\ArrowLine(-40,-30)(-55,-30)

\Gluon(-40,30)(40,-30){3}{15.5} \Gluon(-40,-30)(40,30){3}{15.5}
\BText(0,-50){1l-10}
  \SetWidth{1.6}
\ArrowLine(40,-30)(40,30)
\ArrowLine(40,30)(55,30)
\ArrowLine(55,-30)(40,-30)
\end{picture}}
  }
\]
\vspace*{1.2cm}
\caption{\it One-loop diagrams. Thin arrow lines represent massless quarks, thick arrow line massive quarks, dashed arrow lines are Faddeev-Popov ghosts, and
coiled lines are gluons.}
\label{fig1ltot}
\end{figure}
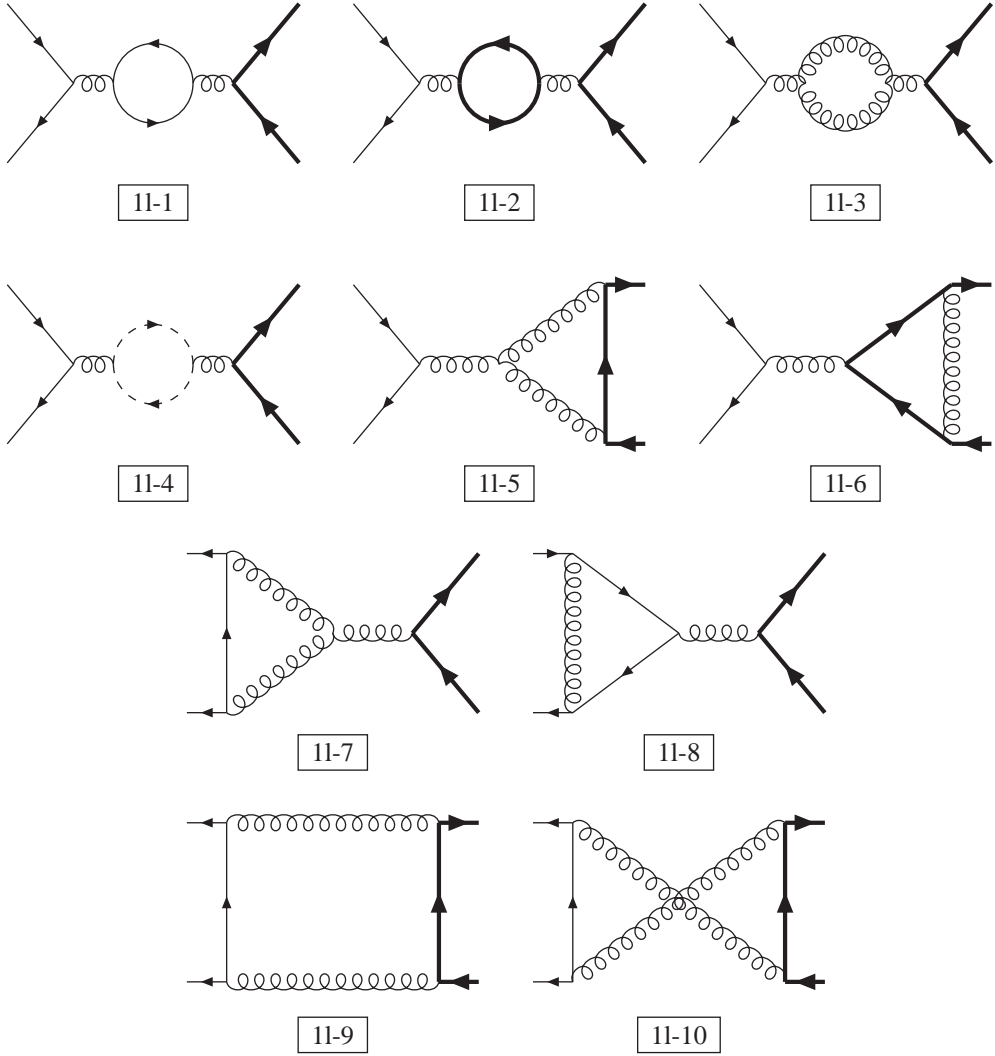

To renormalize the two-loop diagrams contributing  to the $N_l$ corrections of the
partonic cross section it is necessary to extract from the last two lines in
Eq.~(\ref{exprenM}) the terms proportional to $N_l$. Taking into account the fact that
the wave function renormalization factors are zero for the incoming particles and
identical for the massive ones, we find that
\bea
{\mathcal A}^{(2,N_l)} _{{\small \mbox{ren}}} &=&  {\mathcal A}_2^{(N_l)} +
 \left(\delta Z^{(1)}_{{\tiny \mbox{WF}},M}  + 2 \delta
Z^{(1,C_A)}_{\alpha} \right) {\mathcal A}^{(d_1)}_1+
2 \delta Z^{(1,N_l)}_{\alpha} \sum_{j=3}^{10} \nn {\mathcal A}^{(d_j)}_1\\
& & + \left(\delta Z^{(2,N_l)}_{{\tiny \mbox{WF}},M} +
  2 \delta Z^{(1,N_l)}_{\alpha}\delta Z^{(1)}_{{\tiny \mbox{WF}},M}
 +\delta Z^{(2,N_l)}_{\alpha}
\right) {\mathcal A}_0\, .
\label{renNl}
\eea
In Eq.~(\ref{renNl}), the quantity ${\mathcal A}^{(d_j)}_1$ is the amplitude of the
$j$-th diagram in Fig.~\ref{fig1ltot} (stripped of the factor $a$). The
renormalization coefficients not previously defined are:
\bea
\delta Z^{(1,N_l)}_{\alpha} &=& C(\vep) \, \frac{e^{-\gamma \vep}}{\Gamma(1+\vep)} \,
\frac{N_l}{6 \vep} \, , \nn \\
\delta Z^{(1,C_A)}_{\alpha} &=& -C(\vep) \, \frac{e^{-\gamma \vep}}{\Gamma(1+\vep)} \, C_A
\frac{11}{12 \vep}  \, , \nn \\
\delta Z^{(2,N_l)}_{\alpha} &=& C(\vep)^2 \, \left( \frac{e^{-\gamma \vep}}{\Gamma(1+\vep)}
\right)^2 \frac{N_l}{4 \vep} \left(
 \frac{5 }{12}C_A +\frac{1}{4}C_F
   -\frac{11}{9 \vep}C_A \right) \, , \nn \\
\delta Z^{(2,N_l)}_{{\tiny \mbox{WF}},M} &=& C(\vep)^2 \left(\frac{\mu^2}{m^2} \right)^{2 \vep} \, C_F N_l\left(\frac{1}{16 \vep^2} +\frac{9}{32
   \vep} + \frac{59}{64}+\frac{\pi ^2}{24} +
   {\mathcal O} \left( \ep\right)\right) \label{rencoeff} \, .
\eea

In order to renormalize the part of the squared matrix element proportional to $N_h$,
one has to consider the terms proportional to $N_h$ in the last two lines of
Eq.~(\ref{exprenM}), and to add  the counter term for the on-shell mass
renormalization:
\bea
\hspace*{-3mm}
{\mathcal A}^{(2,N_h)} _{{\small \mbox{ren}}}
\hspace*{-1mm} &=& \hspace*{-1mm}  {\mathcal A}_2^{(N_h)} +
 \left(\delta Z^{(1)}_{{\tiny \mbox{WF}},M}  + 2 \delta
Z^{(1,C_A)}_{\alpha} \right) {\mathcal A}^{(d_2)}_1+
2 \delta Z^{(1,N_h)}_{\alpha} \sum_{j=3}^{10} \nn {\mathcal A}^{(d_j)}_1-
2 \delta Z_m^{(1)} {\mathcal A}^{(d_2,\mbox{{\tiny {\tt mass CT }}})}_1\\
\hspace*{-4mm} & & \hspace*{-1mm}   + \left(\delta
Z^{(2,N_h)}_{{\tiny \mbox{WF}},M} + \delta Z^{(2,N_h)}_{{\tiny
\mbox{WF}},m} +
  2 \delta Z^{(1,N_h)}_{\alpha}\delta Z^{(1)}_{{\tiny \mbox{WF}},M}
+ \delta Z^{(2,N_h)}_{\alpha}
\right) {\mathcal A}_0\, .
\label{renNh}
\eea
It must be observed that in this case also the external massless legs acquire a non
vanishing two-loop wave function renormalization factor indicated by $\delta
Z^{(2,N_h)}_{{\tiny \mbox{WF}},m}$. The quantity ${\mathcal A}^{(d_2,\mbox{{\tiny {\tt
mass CT }}})}$ indicates the second diagram in Fig.~\ref{fig1ltot} with a mass counter
term insertion in one of the internal heavy quark lines. The renormalization constant
appearing for the first time in Eq.~(\ref{renNh}) are:
\bea
\delta Z^{(1,N_h)}_{\alpha} &=& C(\vep) \, \frac{e^{-\gamma \vep}}{\Gamma(1+\vep)}  \,
\frac{N_h}{6 \vep} \, , \nn \\
 \delta Z^{(1)}_{m} &=& \delta Z^{(1)}_{{\tiny \mbox{WF}},M}\, , \nn \\
\delta Z^{(2,N_h)}_{\alpha} &=& C(\vep)^2 \, \left( \frac{e^{-\gamma \vep}}{\Gamma(1+\vep)}
\right)^2 \frac{1}{4 \vep} \left(
 \frac{5 }{12}C_A N_h+\frac{1}{4}C_F
   N_h-\frac{11}{9 \vep}C_A N_h\right) \, , \nn \\
\delta Z^{(2,N_h)}_{{\tiny \mbox{WF}},M} &=& C(\vep)^2 \, \left(\frac{\mu^2}{m^2}
\right)^{2 \vep} C_F N_h\left(\frac{1}{8 \vep^2} +\frac{19}{96
   \vep} + \frac{1139}{576}-\frac{\pi ^2}{6}
   +{\mathcal O}\left( \vep\right)\right) \label{rencoeffNh} \, ,
   \nn \\
 \delta Z^{(2,N_h)}_{{\tiny \mbox{WF}},m} &=& C(\vep)^2 \,\left(\frac{\mu^2}{m^2}
\right)^{2 \vep} C_F N_h \left(
 \frac{1}{32 \vep}  -
 \frac{5}{192} + {\mathcal O}\left(\vep \right)
 \right)  \, .
\eea
The renormalization coefficients in Eqs.~(\ref{rencoeff},\ref{rencoeffNh}) can be
found in~\cite{Czakon:2007ej,Melnikov:2000zc}.

The renormalization of the functions $F_i$ ($i=l,h,lh$) in Eq.~(\ref{colstruc}) is
trivial since the relevant two-loop diagrams are reducible and involve the insertion
of two one-loop fermionic vacuum polarization insertions on the gluon propagator in
the diagram of Fig.~\ref{figTree}.

\section{Results\label{sec:results}}

The main result of the present paper is an analytic,
non-approximated  expression for the coefficients $E_l,E_h, D_l,
D_h, F_l, F_{lh}, F_h$ in Eq.~(\ref{colstruc}).  Since such a
result is too long to be explicitly printed here, we included in
the arXiv submission of this work a text file with the complete
result, which is written in terms of one-dimensional HPLs of
maximum weight four  and two-dimensional HPLs of maximum weight
three. Since the coefficients in Eq.~(\ref{colstruc}) still
contain infrared poles, the result is dependent on the choice of
a global, $\vep$-dependent normalization factor. With our
choice, we factor out an overall coefficient
\be
C^2(\vep) = \left[\left(4 \pi \right)^{\vep} \Gamma(1+\vep) \right]^2 \, .
\label{Cep}
\ee
We also provide two codes, one written in {\tt Fortran}, the other as a  {\tt
Mathematica} package, that numerically evaluate  the analytic expression of the
quantities listed above for arbitrary values of the mass scales involved in the
calculation.

In order to cross check our results, we expanded them in the  $s,|t|, |u| \gg m^2$
limit. The first term in the expansion agrees with the results published in
\cite{Czakon:2007ej}; the second order term agrees with the results found in the {\tt
Mathematica} files  included in the arXiv version of \cite{Czakon:2008zk}. We also
find complete agreement with the numerical result of Table~3 in  \cite{Czakon:2008zk},
corresponding to a phase  space point in which the $s,|t|, |u| \gg m^2$ approximation
cannot be applied.

It is straightforward to expand our result for values of the center of mass energy
close to the production threshold. We define
\be
 \beta = \sqrt{1-\frac{4 m^2}{s}} \, , \qquad
 \xi = \frac{1-\cos\theta}{2} \, , \qquad
 L_\mu = \ln\left(\frac{\mu^2}{m^2} \right) \, , \qquad \mbox{ln}_2 = \ln(2) 
\, ,
\ee
where $\theta$ is the scattering angle in the partonic center of mass frame, and we
expand our results in powers of the heavy quark velocity $\beta$, up to terms of order
$\beta^2$.  We find
\begin{eqnarray}
D_l(\beta,\xi) &=& \left(-\frac{1}{4} + {\mathcal O}\left( \beta^2\right) \right)\frac{1}{\vep^3} + \Biggl[
\frac{19}{36}-\frac{\mbox{ln}_2}{3}
+\frac{ L_{\mu}}{6} +\left(-\frac{1}{3}+\frac{2 \xi }{3} \right) \beta +{\mathcal O}\left(\beta ^2\right)\Biggr]\frac{1}{\vep ^2} \nn
\\ \nn &&
+\Biggl\{
2 \mbox{ln}_2^2-\frac{37 \mbox{ln}_2}{9}-\frac{2
   \zeta(2)}{3}+\frac{589}{216}   
   +\left(\frac{37}{18}-2 \mbox{ln}_2\right)  L_{\mu}+\frac{ L_{\mu}^2}{2} 
\\ \nn &&   
   +\left[   
   2-\frac{4 \mbox{ln}_2}{3}+\left(\frac{8 \mbox{ln}_2}{3}-4\right) \xi    
   +\left(\frac{2}{3}-\frac{4 \xi }{3} \right)  L_{\mu} \right] \beta +{\mathcal O}\left(\beta ^2\right)\Biggr\}\frac{1}{\vep }
\\ \nn &&   
   +
\Biggl\{
-\frac{32 \mbox{ln}_2^3}{9}+\frac{16
   \mbox{ln}_2^2}{9}+\frac{14 \zeta(2) \mbox{ln}_2}{3}+\frac{475 \mbox{ln}_2}{54}-\frac{13 \zeta(2)}{18}-\frac{79
   \zeta(3)}{18}-\frac{1211}{144}
\\ \nn &&   
   +\left(\frac{16 \mbox{ln}_2^2}{3}-\frac{4 \mbox{ln}_2}{3}-\frac{7 \zeta(2)}{3}-\frac{163}{36}\right)
    L_{\mu}+\left(\frac{1}{9}-\frac{8 \mbox{ln}_2}{3}\right)  L_{\mu}^2+\frac{4  L_{\mu}^3}{9} 
\\ \nn &&        
    +\Biggl[\frac{20 \mbox{ln}_2^2}{9}-\frac{64
   \mbox{ln}_2}{27}-\frac{26 \zeta(2)}{9}+\frac{7}{27}+\left(-\frac{40 \mbox{ln}_2^2}{9}+\frac{128 \mbox{ln}_2}{27}+\frac{52
   \zeta(2)}{9}-\frac{14}{27}\right) \xi  
\\  &&   
   +\!\left(\!\frac{20}{9}\!-\!\frac{20 \mbox{ln}_2}{9}\!+\!\left(\frac{40
   \mbox{ln}_2}{9}\!-\!\frac{40}{9}\right)\! \xi \! \right)  L_{\mu}\!+\!\left(\frac{4}{3}\!-\!\frac{8 \xi }{3} \right)
    L_{\mu}^2 \Biggr] \beta \!+\!{\mathcal O}\left(\beta ^2\right)\!\Biggr\}\!+\!
    {\mathcal O}\left(\vep\right) \, ,\\ \nn
D_h(\beta,\xi) & = & \left(\frac{8}{9}+\frac{2  L_{\mu}}{3} +{\mathcal O}\left(\beta ^2\right)\right)\frac{1}{\vep ^2}
+\Biggl\{-\frac{16   \mbox{ln}_2}{9}+\frac{\zeta(2)}{3}+\frac{88}{27}+\left(\frac{25}{9}-\frac{4 \mbox{ln}_2}{3}\right)
    L_{\mu}+ L_{\mu}^2 
\\ \nn &&    
    +\!\left[\frac{16}{9}-3 \zeta(2)-\frac{32 \xi }{9}+\left(\frac{4}{3}-\frac{8 \xi }{3} \right)  L_{\mu} \right] \beta +{\mathcal O}\left(\beta
   ^2\right)\Biggr\}\frac{1}{\vep }
   +\!\Biggl\{
   \frac{16 \mbox{ln}_2^2}{9}+\frac{55 \zeta(2) \mbox{ln}_2}{3}
\\ \nn &&   
-\frac{148 \mbox{ln}_2}{27}-\frac{857
   \zeta(2)}{72}+\frac{283 \zeta(3)}{144}-\frac{209}{108}+\left(\frac{4 \mbox{ln}_2^2}{3}-\frac{14
   \mbox{ln}_2}{9}-\frac{\zeta(2)}{3}-\frac{319}{54}\right)  L_{\mu}
\\ \nn  &&   
   +\left(\frac{5}{6}-2 \mbox{ln}_2\right)  L_{\mu}^2+\frac{7
    L_{\mu}^3}{9} 
    +\Biggl[12 \zeta(2) \mbox{ln}_2+\frac{8 \mbox{ln}_2}{9}-\frac{131 \zeta(2)}{18}+6 \zeta(2) \ln (\beta
   )
\\ \nn &&   
   +\frac{214}{27}   
   +\left(-\frac{16 \mbox{ln}_2}{9}-\frac{58 \zeta(2)}{9}-\frac{428}{27}\right) \xi 
   +\Biggl(\frac{4 \mbox{ln}_2}{9}-6 \zeta(2)+\frac{16}{9}
\\  &&   
   +\left(-\frac{8 \mbox{ln}_2}{9}-\frac{32}{9}\right) \xi
    \Biggr)  L_{\mu}+\left(2-4 \xi  \right)  L_{\mu}^2 \Biggr] \beta +{\mathcal O}\left(\beta
   ^2\right)\Biggr\}+{\mathcal O}\left(\vep \right) \, ,\\ 
\nn 
E_l(\beta,\xi) & = & \left(\frac{1}{4} + {\mathcal O}\left(\beta
   ^2\right) \right)\frac{1}{ \vep^3}
+\left[\frac{\mbox{ln}_2}{3}-\frac{25}{36}-\frac{ L_{\mu}}{6}
 +\left(\frac{4}{3}-\frac{8 \xi
   }{3} \right) \beta +{\mathcal O}\left(\beta ^2\right)\right]\frac{1}{\vep ^2}
\\ \nn &&   
   +\Biggl\{\frac{\zeta(2)}{\beta }-2 \mbox{ln}_2^2+\frac{31
   \mbox{ln}_2}{9}+\frac{8 \zeta(2)}{3}-\frac{373}{216}+\left(2 \mbox{ln}_2-\frac{31}{18}\right)
    L_{\mu}-\frac{ L_{\mu}^2}{2} 
\\ \nn &&    
    +\Biggl[\frac{16 \mbox{ln}_2}{3}+\zeta(2)-8+\left(-\frac{32 \mbox{ln}_2}{3}-2
   \zeta(2)+16\right) \xi +2 \zeta(2) \xi ^2 
\\ \nn &&   
   +\left(-\frac{8}{3}+\frac{16 \xi }{3} \right)
    L_{\mu} \Biggr] \beta +{\mathcal O}\left(\beta ^2\right)\Biggr\}\frac{1}{\vep }
    +\Biggl\{\Biggl(-8 \mbox{ln}_2 \zeta(2)-4 \ln (\beta ) \zeta(2)+4 \zeta(2)
\\ \nn &&    
    +4
   \zeta(2)  L_{\mu} \Biggr)\frac{1}{\beta }+
   \frac{32 \mbox{ln}_2^3}{9}-\frac{68 \mbox{ln}_2^2}{9}-\frac{50 \zeta(2) \mbox{ln}_2}{3}+\frac{643
   \mbox{ln}_2}{54}+\frac{221 \zeta(2)}{18}
   -\frac{10 \zeta(3)}{9}
\\ \nn && 
-\frac{10285}{1296}+\left(-\frac{16 \mbox{ln}_2^2}{3}+\frac{68
   \mbox{ln}_2}{9}+\frac{25 \zeta(2)}{3}-\frac{787}{108}\right)  L_{\mu}+\left(\frac{8 \mbox{ln}_2}{3}-\frac{17}{9}\right)  L_{\mu}^2-\frac{4
    L_{\mu}^3}{9} 
\\ \nn &&    
    +\Biggl[
    -\frac{80 \mbox{ln}_2^2}{9}-8 \zeta(2) \mbox{ln}_2+\frac{256 \mbox{ln}_2}{27}+\frac{158
   \zeta(2)}{9}-4 \zeta(2) \ln (\beta )-\frac{28}{27}
\\ \nn &&   
   +\left(\frac{160 \mbox{ln}_2^2}{9}+16 \zeta(2) \mbox{ln}_2-\frac{512
   \mbox{ln}_2}{27}-\frac{334 \zeta(2)}{9}+8 \zeta(2) \ln (\beta )+\frac{56}{27}\right) \xi
\\ \nn &&   
    +\Bigl(-16 \mbox{ln}_2 \zeta(2)-8 \ln (\beta ) \zeta(2)+14
   \zeta(2)\Bigr) \xi ^2  
   +\Biggl(
\frac{80 \mbox{ln}_2}{9}+4 \zeta(2)-\frac{80}{9}
   +\Bigl(-\frac{160 \mbox{ln}_2}{9}
\\ &&   -8
   \zeta(2)+\frac{160}{9}\Bigr) \xi \!+\!8 \zeta(2) \xi ^2 \Biggr)  L_{\mu}\!+\!\left(-\frac{16}{3}\!+\!\frac{32 \xi }{3}\right)  L_{\mu}^2 \Biggr] \beta \!+\!{\mathcal O}\left(\beta ^2\right)\Biggr\}\!+\!{\mathcal O}\left(\vep\right)
   \, , \\ 
\nn 
E_h(\beta,\xi) & = & \left(-\frac{8}{9}-\frac{2  L_{\mu}}{3} +{\mathcal O}\left(\beta ^2\right)\right)\frac{1}{\vep ^2}+\Biggl\{
\frac{16 \mbox{ln}_2}{9}+\frac{7
   \zeta(2)}{6}-\frac{64}{27}+\left(\frac{4 \mbox{ln}_2}{3}-\frac{19}{9}\right)  L_{\mu}- L_{\mu}^2 
\\ \nn &&   
   +\left[6
   \zeta(2)-\frac{64}{9}+\frac{128 \xi }{9} +\left(-\frac{16}{3}+\frac{32 \xi }{3} \right)
    L_{\mu} \right] \beta +{\mathcal O}\left(\beta ^2\right)\Biggr\}
    \frac{1}{\vep }
\\ \nn &&
    +\Biggl\{\left(\frac{8 \zeta(2)}{3}+2 \zeta(2)  L_{\mu} \right)\frac{1}{\beta
   }
   -\frac{16 \mbox{ln}_2^2}{9}-\frac{43 \zeta(2) \mbox{ln}_2}{3}+\frac{343 \mbox{ln}_2}{27}+\frac{61 \zeta(2)}{9 \sqrt{2}}
\\ \nn &&   
   +\frac{841
   \zeta(2)}{72}-\frac{83 \zeta(3)}{18}-12 \zeta(2) \ln (\beta )-\frac{5}{12} \mbox{Li}_3\left(3-2 \sqrt{2}\right)-\frac{61 \mbox{Li}_2\left(3-2
   \sqrt{2}\right)}{9 \sqrt{2}}
\\ \nn &&   
   +\frac{5}{18} \ln ^3\left(1+\sqrt{2}\right)-\frac{61 \ln ^2\left(1+\sqrt{2}\right)}{9 \sqrt{2}}-\frac{5}{6} \zeta(2) \ln
   \left(1+\sqrt{2}\right)-\frac{6703}{324}
\\ \nn &&   
   +\left(-\frac{4 \mbox{ln}_2^2}{3}+\frac{38 \mbox{ln}_2}{9}+\frac{16 \zeta(2)}{3}-\frac{463}{54}\right)
    L_{\mu}+\left(2 \mbox{ln}_2-\frac{41}{18}\right)  L_{\mu}^2-\frac{7  L_{\mu}^3}{9} 
\\ \nn  &&    
    +\Biggl[
    -24 \zeta(2) \mbox{ln}_2-\frac{32
   \mbox{ln}_2}{9}+\frac{7 \zeta(2)}{9}-12 \zeta(2) \ln (\beta )-\frac{856}{27}+\Biggl(\frac{64 \mbox{ln}_2}{9}+\frac{184
   \zeta(2)}{9}
\\ \nn &&   
   +\frac{1712}{27}\Biggr) \xi +\frac{16 \zeta(2) \xi ^2}{3} +\Biggl(-\frac{16 \mbox{ln}_2}{9}+14
   \zeta(2)-\frac{64}{9}+\left(\frac{32 \mbox{ln}_2}{9}-4 \zeta(2)+\frac{128}{9}\right) \xi 
\\ &&  
   +4 \zeta(2) \xi ^2 \Biggr)
    L_{\mu}+\left(-8+16 \xi  \right)  L_{\mu}^2 \Biggr] \beta +{\mathcal O}\left(\beta ^2\right)\Biggr\}+{\mathcal O}\left(\vep\right) \, , \\ 
F_l(\beta,\xi) & = & \frac{2}{81} \left[(5-6 \mbox{ln}_2)^2-54 \zeta(2)\right]+{\mathcal O}\left(\beta ^2\right)+{\mathcal O}\left(\vep\right) \, , \\ 
F_h(\beta,\xi) &=& \frac{128}{81}+{\mathcal O}\left(\beta ^2\right)+{\mathcal O}\left(\vep\right) \, , \\ 
F_{lh}(\beta,\xi) &=& \left(\frac{160}{81}-\frac{64 \mbox{ln}_2}{27}\right)-4 \zeta(2) \beta +{\mathcal O}\left(\beta ^2\right)+{\mathcal O}\left(\vep\right)\, . 
\end{eqnarray}
These
expansions could be used in the future in the calculation of
logarithmically enhanced terms at production  threshold.

\section{Conclusions and Outlook\label{sec:conc}}

In this paper, we presented the analytic calculation of the two-loop fermionic
corrections to the heavy-quark production amplitude for $q\bar q \to t\bar t$,
retaining the exact heavy-quark mass dependence. Our work serves as an independent
confirmation of recent results obtained analytically as small-mass
expansions~\cite{Czakon:2007ej} and numerically~\cite{Czakon:2008zk}. We also provide
new results on the threshold expansion of the amplitude.

Our result represents a gauge invariant sub-set of the full two-loop virtual
correction to the partonic process $q \overline{q} \to t \overline{t}$. In order to
complete the analytic calculation  of the two-loop virtual corrections, it is
necessary to calculate the diagrams that do not contain fermion
loops, which is currently in progress. Likewise, analytic
results for the two-loop amplitude for $gg\to t\bar t$ could be obtained in the same
calculational framework.

In order to obtain NNLO predictions for the total $t \overline{t}$
production cross section and for differential distributions, it is
necessary to combine the two-loop virtual corrections with the
already available~\cite{Dittmaier:2007wz} one-loop corrections to
the $t\bar t$+(1~parton) process and the tree-level  $t\bar
t$+(2~partons) process. Since these contribute to
infrared-divergent configurations where up to two partons can
become unresolved, their implementation requires the application
of a NNLO subtraction method. The methods presently
available~\cite{secdec,ant,cg} have been applied up to
now~\cite{babis,our3j,cghiggs} to at most $1\to 3$ processes in
$e^+e^-$ annihilation and $2\to 1$ processes at hadron colliders,
such that a calculation of a hadronic $2\to 2$ process, involving
massive partons, will be a new step in complexity, potentially
requiring further refinements of these methods.

\subsection*{Acknowledgments}

R.~B.\ would like to thank S.~Catani for useful discussions. We
are grateful to J.~Vermaseren  for his kind assistance in the use
of {\tt FORM} \cite{FORM}. This work  was supported  by the Swiss
National Science Foundation (SNF) under contracts 200020-117602 and 
PBZH2-117028.
The work of D.~M.  was partly supported by the US Department of
Energy under contract DE-AC02-76SF00515.
%

\appendix

\section{Harmonic Polylogarithms \label{AppGPLs}}

The results of this work are conveniently expressed, in the
non-physical region $s < 0$, in terms of one- and two-dimensional
HPLs. Nowadays, harmonic polylogarithms are extensively used in
multiloop computations, therefore in this section we just
summarize their definition; the reader interested in the algebraic
properties of these functions can  find detailed discussions of
the topic in the available literature \cite{HPLs,2dHPLs}.

In the process under study, five different weight functions are needed; they are
\be
f_w(x) = \frac{1}{x-w}\, ,\quad \mbox{with} \quad w \in \left\{ 0,1,-1,-y,-\frac{1}{y} \right\}
\, .
\ee
The weight-one HPLs are defined as
\be
G(0;x) = \ln{x} \, , \qquad G(w;x) = \int_0^x dt f_w(t) \, .
\ee
 HPLs of higher weight are defined by iterated  integrations
\be
G(w,\cdots;x) = \int_0^x dt f_w(t) G(\cdots;t) \, ,
\ee
with the only exception of the HPLs in which all the weights are zero which are
defined as follows
\be
G(\underbrace{0,0,\cdots,0}_n;x) = \frac{1}{n!} \ln^n{x} \,.
\ee
The reader should be aware of the fact that in the original definition of Remiddi and
Vermaseren, the weight function corresponding to the weight $1$ was $f_1 =1/(1-x)$. In
order to translate the HPLs defined with the Remiddi-Vermaseren convention to the ones
employed in this work (and vice versa) it is sufficient to multiply each HPL by a
factor $(-1)^n$, where $n$ is the number of weights equal to $1$.

The weights $-y$ and $-1/y$ were already introduced in \cite{2dHPLs,electronloop}. In
our results, the two-dimensional harmonic polylogarithms have maximum weight three. As
it is well known, if the weight is not larger than three, these functions can be
rewritten in terms of Nielsen polylogarithms. For completeness, we list below the
explicit expression of the two-dimensional harmonic polylogarithms which appear in
our analytic results in terms of Nielsen's polylogarithms:
\bea
G(-y;x) &=& \ln\left(\frac{x+y}{y} \right) \, , \nn \\
G(-1/y;x) &=&  \ln\left(1+ x y \right)\, , \nn \\
G(-y,0;x) &=& \ln (x) \ln \left(\frac{x+y}{y}\right)+\mbox{Li}_2\left(-\frac{x}{y}\right)\, , \nn \\
G(-1/y,0;x) &=& \ln (x) \ln (x y+1)+\mbox{Li}_2(-x y) \, , \nn \\
G(-y,1;x) &=& \frac{1}{2} \ln ^2(y+1)-\ln (1-x) \ln (y+1)-\ln (y) \ln (y+1)
\nn \\ &&+\ln (1-x) \ln
  (x+y)-\mbox{Li}_2(-y)+\mbox{Li}_2\left(\frac{1-x}{y+1}\right)-
  \frac{\pi ^2}{6}\, , \nn \\
G(-1/y,1;x) &=& \frac{1}{2} \ln ^2(y+1)-\ln (1-x) \ln (y+1)+\ln (1-x)
\ln (x y+1)\nn \\ &&
+\mbox{Li}_2(-y)+\mbox{Li}_2\left(\frac{y-x y}{y+1}\right) \, , \nn \\
G(-y,0,0;x) &=&\frac{1}{2} \ln^2(x)\ln \left(1+\frac{x}{y} \right)  + \ln(x) \mbox{Li}_2
\left(- \frac{x}{y}\right) - \mbox{Li}_3 \left(-\frac{x}{y}\right)\, , \nn \\
G(-1/y,0,0;x) &=& \frac{1}{2} \ln^2(x)\ln \left(1+x y \right)  + \ln(x) \mbox{Li}_2
\left(- xy\right) - \mbox{Li}_3 \left(-xy\right) \, , \nn \\
G(-y,1,0;x) &=& -\frac{1}{3} \ln ^3(1\!-\!x)\!-\!\ln (x) \ln
^2(1\!-\!x)\!-\!\ln (y) \ln
   ^2(1\!-\!x)+\frac{1}{2} \ln (y\!+\!1) \ln ^2(1\!-\!x)
\nn \\ &&
   +\ln (x+y) \ln ^2(1-x)-\frac{1}{2}
   \ln ^2(y+1) \ln (1-x)-\ln ^2(x+y) \ln (1-x)
\nn \\ &&
    -\!\ln (x) \ln (y) \ln (1\!-\!x)\!-\!\ln (x)
   \ln (y\!+\!1) \ln (1\!-\!x)\!+\!\ln (y) \ln (y\!+\!1) \ln (1\!-\!x)
\nn \\ &&
+2 \ln (x) \ln (x+y) \ln
   (1-x)+\ln (y) \ln (x+y) \ln (1-x)
\nn \\ &&
   -\mbox{Li}_2(x) \ln
   (1-x)+\mbox{Li}_2\left(-\frac{x}{y}\right) \ln (1-x)
    +\mbox{Li}_2(-y) \ln(1-x)
\nn \\ &&
    -\mbox{Li}_2\left(\frac{(1-x) y}{x+y}\right) \ln
   (1-x)-\mbox{Li}_2\left(\frac{x+y}{x-1}\right) \ln (1-x)-\frac{1}{3} \pi ^2
   \ln (1-x)
\nn \\ &&
    +\frac{1}{2} \ln (x) \ln ^2(y+1)-\frac{1}{6} \pi ^2 \ln (x)-\ln (x)
   \ln (y) \ln (y+1)-\ln (y) \mbox{Li}_2(x)
\nn \\ &&
    +\ln (x\!+\!y) \mbox{Li}_2(x)\!-\!\ln (x)
   \mbox{Li}_2(-y)+\ln (x) \mbox{Li}_2\left(\frac{1-x}{y+1}\right)+\ln (y)
   \mbox{Li}_2\left(\!\frac{y}{x\!+\!y}\!\right)
\nn \\ &&
    -\ln (x\!+\!y)
   \mbox{Li}_2\left(\frac{y}{x\!+\!y}\right)\!+\!\ln (y) \mbox{Li}_2\left(\frac{(1\!-\!x)
   y}{x+y}\right)+\ln (x\!+\!y) \mbox{Li}_2\left(\!\frac{(1\!-\!x) y}{x\!+\!y}\!\right)
\nn \\ &&
   -\!\ln (y)
   \mbox{Li}_2\left(\frac{x\!+\!y}{x-1}\right)+ \ln (x\!+\!y)    \mbox{Li}_2
   \left(\!\frac{x\!+\!y}{x\!-\!1}\!\right)- \mbox{Li}_3(1-x)-
   \mbox{Li}_3\left(\!\frac{x}{x\!-\!1}\!\right)
\nn \\ &&
   -\mbox{Li}_3
   \left(-\frac{x}{y}\right)-\mbox{Li}_3(-y)
   +\mbox{Li}_3\left(\frac{1}{y+1}\right)-
   \mbox{Li}_3\left(\frac{1-x}{y+1}\right)
\nn \\ &&
    +\mbox{Li}_3\left(\frac{x (y+1)}{(x\!-\!1)
  y}\right)\!-\!\mbox{Li}_3\left(\frac{y}{x\!+\!y}\right)\!+
  \mbox{Li}_3\left(\frac{(1\!-\!x)
   y}{x\!+\!y}\right)\!-\!\mbox{Li}_3\left(\frac{x\!+\!y}{x\!-\!1}\right)+\zeta(3) \, , \nn \\
G(-1/y,1,0;x) &=& -\frac{1}{3} \ln ^3(1-x)-\ln (x) \ln ^2(1-x)
-\frac{1}{2} \ln
   (y) \ln ^2(1-x)
\nn \\ &&
+\frac{1}{2} \ln (y+1) \ln ^2(1-x)+\ln (x
   y+1) \ln ^2(1-x)-\frac{1}{2} \ln ^2(y+1) \ln (1-x)
\nn \\ &&
-\ln^2(x y+1) \ln (1-x)-\ln (x) \ln (y+1) \ln (1-x)
\nn \\ &&
+\!2 \ln
   (x) \ln (x y\!+\!1) \ln (1\!-\!x)
\!+\!\ln (y) \ln (x y\!+\!1) \ln
   (1\!-\!x)-\mbox{Li}_2(x) \ln (1\!-\!x)
\nn \\ &&
-\mbox{Li}_2(-y) \ln
   (1-x)+\mbox{Li}_2(-x y) \ln
   (1-x)-\mbox{Li}_2\left(\frac{1-x}{x y+1}\right) \ln
   (1-x)
\nn \\ &&
-\mbox{Li}_2\left(\frac{x y+1}{(x-1) y}\right) \ln
   (1-x)+\frac{1}{6} \pi ^2 \ln (1-x)+\frac{1}{2} \ln (x) \ln
   ^2(y+1)
\nn \\ &&
+\ln (x y+1) \mbox{Li}_2(x)+\ln (x)
   \mbox{Li}_2(-y)
+\ln (x) \mbox{Li}_2\left(\frac{y-x
   y}{y+1}\right)
\nn \\ &&
-\ln (x y+1) \mbox{Li}_2\left(\frac{1}{x
   y+1}\right)+\ln (x y+1) \mbox{Li}_2\left(\frac{1-x}{x
   y+1}\right)
\nn \\ &&
+\ln (x y+1) \mbox{Li}_2\left(\frac{x y+1}{(x-1)
   y}\right)-\mbox{Li}_3 \left(1-x\right)-\mbox{Li}_3\left(\frac{x}{x-1}
   \right)+\mbox{Li}_3\left(-\frac{1}{y}\right)
\nn \\ &&
-\mbox{Li}_3(-x
   y)+\mbox{Li}_3\left(\frac{y}{y+1}\right)+\mbox{Li}_3\left(
   \frac{x (y+1)}{x-1}\right)-\mbox{Li}_3\left(\frac{y-x
   y}{y+1}\right)
\nn \\ &&
-\mbox{Li}_3\left(\frac{1}{x
   y+1}\right)+\mbox{Li}_3\left(\frac{1-x}{x
   y+1}\right)-\mbox{Li}_3\left(\frac{x y+1}{(x-1)
   y}\right)+\zeta (3) \, , \nn \\
G(-y,0,1;x) &=& \frac{1}{3} \ln ^3(1\!-\!x)+\ln (x) \ln
^2(1\!-\!x)+\frac{1}{2} \ln
   (y) \ln ^2(1\!-\!x)\!-\!\ln (x\!+\!y) \ln ^2(1\!-\!x)
\nn \\ &&
    +\ln ^2(x+y) \ln (1-x)+\ln (x)
   \ln (y) \ln (1-x)
    -\ln (x) \ln (x+y) \ln (1-x)
\nn \\ &&
    -\ln (y) \ln (x\!+\!y) \ln
   (1\!-\!x)+\mbox{Li}_2(x) \ln (1\!-\!x)+\mbox{Li}_2\left(\!\frac{(1-x) y}{x+y}\!\right)
   \ln (1-x)
\nn \\ &&
    +\mbox{Li}_2\left(\frac{x+y}{x-1}\right) \ln (1-x)-\frac{1}{6} \pi
   ^2 \ln (1-x)+\ln (y) \mbox{Li}_2(x)
\nn \\ &&
    -\ln (x+y) \mbox{Li}_2(x)-\ln (y)
   \mbox{Li}_2\left(\frac{y}{x+y}\right)+\ln (x+y)
   \mbox{Li}_2\left(\frac{y}{x+y}\right)
\nn \\ &&
    +\ln (y) \mbox{Li}_2\left(\frac{(1-x)
   y}{x+y}\right)-\ln (x+y) \mbox{Li}_2\left(\frac{(1-x) y}{x+y}\right)+\ln
   (y) \mbox{Li}_2\left(\frac{x+y}{x-1}\right)
\nn \\ &&
    - \ln (x+y)
   \mbox{Li}_2\left(\frac{x+y}{x-1}\right)+\mbox{Li}_3(1-x)
   -\mbox{Li}_3(-y)+\mbox{Li}_3\left(\frac{y}{x+y}\right)
\nn \\ &&
   -\mbox{Li}_3\left(\frac{(1-x)
   y}{x+y}\right)
    +\mbox{Li}_3\left(\frac{x+y}{x-1}\right)-\zeta(3) \, , \nn \\
G(-1/y,0,1;x) &=& \frac{1}{3} \ln ^3(1-x)+\ln (x) \ln ^2(1-x)+\frac{1}{2}
   \ln (y) \ln ^2(1-x)
\nn \\ &&-\ln (x y+1) \ln ^2(1-x)
    +\ln ^2(x y+1) \ln
   (1-x)-\ln (x) \ln (x y+1) \times
\nn \\ && \times \ln (1-x)
    -\ln (y) \ln (x y+1) \ln
   (1-x)
    +\mbox{Li}_2\left(\frac{1-x}{x y+1}\right) \ln
   (1-x)
\nn \\ &&
    +\mbox{Li}_2\left(\frac{x y+1}{(x-1) y}\right) \ln (1-x)-\frac{1}{6}
   \pi ^2 \ln (1-x)+\ln \left(\frac{1-x}{x y+1}\right) \mbox{Li}_2(x)
\nn \\ &&
    +\ln (x
   y+1) \mbox{Li}_2\left(\frac{1}{x y+1}\right)-\ln (x y+1)
   \mbox{Li}_2\left(\frac{1-x}{x y+1}\right)
\nn \\ &&
    -\ln (x y+1)
   \mbox{Li}_2\left(\frac{x y+1}{(x-1) y}\right)
   +\mbox{Li}_3(1-x)-\mbox{Li}_3\left(-\frac{1}{y}\right)
\nn \\ &&
   +\mbox{Li}_3\left(\frac{1}{x y+1}\right)-\mbox{Li}_3\left(\frac{1-x}{x
   y+1}\right)+\mbox{Li}_3\left(\frac{x y+1}{(x-1) y}\right)-\zeta(3) \, , \nn
   \\
G(-y,1,1;x) &=& -\frac{1}{2} \ln (y+1) \ln ^2(1-x)+\frac{1}{2} \ln (x+y) \ln ^2(1-x)
\nn \\ &&
+\mbox{Li}_2\left(\frac{1-x}{y+1}\right)
\ln(1-x)+\mbox{Li}_3\left(\frac{1}{y+1}\right)-
\mbox{Li}_3\left(\frac{1-x}{y+1}\right) \, , \nn \\
G(-1/y,1,1;x) &=& -\frac{1}{2} \ln (y+1) \ln ^2(1-x)+\frac{1}{2} \ln (x y+1)
\ln ^2(1-x)\nn \\ &&+\mbox{Li}_2\left(\frac{y-x y}{y+1}\right)
\ln(1-x)+\mbox{Li}_3\left(\frac{y}{y+1}\right)-
\mbox{Li}_3\left(\frac{y-x y}{y+1}\right) \, .
\eea

We first obtained the squared matrix elements in the  non-physical region $s < 0$. The
corresponding quantities in the  physical region $s >0$  could be obtained by analytic
continuation to the complex value $s \to s + i \delta$, where $\delta \to 0^+$. For $s
> 4 m^2$ the variable $x$ becomes
\be
x = -x' + i \delta \, ,
\ee
where
\be
x' = \frac{\sqrt{s}-\sqrt{s-4 m^2}}{\sqrt{s} + \sqrt{s-4 m^2}} \, ,
\ee
So that $0<x'<1$ for $4 m^2<s<\infty$. While the HPLs with argument $y$ and $z$ are
always real, the HPLs of $x$ are complex for $s > 0$. In particular, the imaginary
part of the HPLs of argument $x$ for $s > 4 m^2$ is defined when the analytic
continuation of the logarithm is specified:
\be \label{ancon}
G(0;x) \to G(0;-x'+ i \delta) = G(0,x') + i \pi\, .
\ee
In the quantities $E_h$ and $D_h$ one also encounters  the
variable $x_p$ defined in Eq.~(\ref{xp}). In this case, in the
physical region one finds that
\be
 x_p \to -x'_p + i \delta \, ,
\ee
where
\be
x'_p = \frac{\sqrt{s+4 m^2}-\sqrt{s}}{\sqrt{s+4 m^2} + \sqrt{s}} \, ;
\ee
therefore $ 0< x_p <1$ when $0<s<\infty$. The imaginary part of the HPLs of argument
$x'_p$ is defined according to the same principle that determines the imaginary part
of the HPLs of argument $x$ (Eq.(\ref{ancon})). The analytic continuation of the
two-dimensional HPLs presented in this appendix is determined by the analytic
properties of the functions $\ln$, $\mbox{Li}_2$ and   $\mbox{Li}_3$.

\section{Master Integrals \label{AppMIs}}

In this Appendix we collect the MIs for the topologies in Fig.~\ref{nlMIs}-(k),
Fig.~\ref{nlMIs}-(l),   Fig.~\ref{nhMIs}-(k), and  in Fig.~\ref{nhMIs}-(l)  that are
not yet available in the literature.

The explicit expression of the MIs depends on the chosen normalization of the
integration measure. The integration on the loop momenta is normalized as follows
\be
\int{\mathfrak
D}^dk = \frac{1}{C(\vep)} \left( \frac{\mu^2}{m^2}
\right)^{-\vep} \int \frac{d^d k}{(4 \pi^2)^{(1-\vep)}} \, ,
\label{measure}
\ee
where $C(\vep)$ was defined in Eq.~(\ref{Cep}). In Eq.~(\ref{measure}) $\mu$ stands
for the 't Hooft mass of dimensional regularization. The integration measure in
Eq.~(\ref{measure}) is chosen in such a way that the one-loop massive tadpole becomes
\be
\int{\mathfrak D}^dk \ \frac{1}{k^2+m^2} =
              -\frac {m^2}{4 (1- \vep) \vep} \, .
\label{Tadpole}
\ee
In calculating the squared matrix element, we multiply our bare results by
$(\mu^2/m^2)^\vep$, in order to make explicit the dependence on the top scale. We also
point out that, since the squared matrix element still contains soft and collinear
divergencies regulated by $\ep$, it depends on the normalization of the integration
measure. In particular, in order to match our results with the ones of
\cite{Czakon:2007ej,Czakon:2008zk}, it is necessary to multiply them by the factor
\be
\frac{e^{-2 \gamma_{\mbox{{\tiny}}} \vep}}{\Gamma\left(1+\vep \right)^2} =
1- \zeta(2) \vep^2 + \frac{2}{3} \zeta(3) \vep^3 + {\mathcal O}\left(
 \vep^4\right) \, .
\ee

There is a single MIs belonging to the topology Fig.~\ref{nlMIs}-(k) :

\begin{eqnarray}
\hbox{\begin{picture}(0,0)(0,0)
\SetScale{1}
  \SetWidth{.5}
  \Line(-30,-20)(30,-20)
  \Line(-20,20)(20,-20)
  \Line(-20,20)(30,20)
  \Line(-30,20)(-20,20)
\CArc(0,42.5)(30,228,312)
  \SetWidth{1.6}
  \Line(20,20)(30,20)
  \Line(20,-20)(30,-20)
  \Line(20,20)(20,-20)
\end{picture}}
& \hspace*{10mm} = & \int \frac{{\mathfrak
D}^dk_1  {\mathfrak
D}^dk_2}{P_0\left(k_1 + p_1\right) P_0\left(k_2\right)
P_0 \left(k_1-k_2\right) P_m\left(k_1+p_3\right)} \, ,
\end{eqnarray}

\vspace*{5mm}
\noindent where we define
\be
P_0\left(k\right) \equiv k^2 \, , \qquad P_m\left(k\right) \equiv k^2 +m^2 \, ;
\ee
we then find

\begin{eqnarray}
\hbox{\begin{picture}(0,0)(0,0)
\SetScale{1}
  \SetWidth{.5}
  \Line(-30,-20)(30,-20)
  \Line(-20,20)(20,-20)
  \Line(-20,20)(30,20)
  \Line(-30,20)(-20,20)
\CArc(0,42.5)(30,228,312)
  \SetWidth{1.6}
  \Line(20,20)(30,20)
  \Line(20,-20)(30,-20)
  \Line(20,20)(20,-20)
\end{picture}}
& \hspace*{16mm} = & \sum_{i=-2}^1 A_i \, \vep^i + {\mathcal O}(\vep^2) \, ,
\end{eqnarray}

\begin{eqnarray} \nonumber
 A_{-2} &=& \frac{1}{32}\,, \\ \nonumber
 A_{-1} &=&  \frac{1}{32 y}\left[5 y - 2  \left(1 + y\right) G \left(-1;y\right)\right] \,,\\ \nonumber
 A_{0} &=&
 \frac{1}{32 y} \left[19 y + 4 y \zeta(2) - 10  \left(1 + y\right)
     G \left(-1;y\right) \right. \\ \nn
     && \left.+ 8  \left(1 + y\right) G \left(-1,-1;y\right) - 4
     G \left(0,-1;y\right) - 4 y G \left(0,-1;y\right)\right]  \, ,
     \\ \nonumber
 A_{1} &=&  \frac{1}{32 y}
 \Bigl[65 y + 20 y
      \zeta \left(2\right) + 8 y  \zeta \left(3\right) - 2  \left(1 + y\right)  \left(19 + 4  \zeta \left(2\right)\right)
      G \left(-1;y\right)
      \\ \nn &&  + 40  \left(1 + y\right)  G \left(-1,-1;y\right) - 20  G \left(0,-1,y\right) - 20 y  G \left(0,-1;y\right)
      - 32  G \left(-1,-1,-1;y\right)
      \\ \nn &&  - 32 y  G \left(-1,-1,-1;y\right) + 16  G \left(-1,0,-1,y\right) + 16 y
      G \left(-1,0,-1;y\right)
      \\  &&
      +\! 16  G \left(0,-1,-1;y\right)
       \!+\! 16 y  G \left(0,-1,-1;y\right) \!-\! 8
      G \left(0,0,-1;y\right) \!- \!`8 y  G \left(0,0,-1;y\right)\!\Bigr]\, .
\end{eqnarray}

Also the four point topology Fig.~\ref{nlMIs}-(l) has a single MI:

\begin{eqnarray}
\hbox{\begin{picture}(0,0)(0,0)
\SetScale{1}
  \SetWidth{.5}
  \Line(-30,20)(30,20)
  \Line(-30,-20)(30,-20)
  \Line(-20,20)(-20,-20)
\CArc(0,42.5)(30,228,312)
  \SetWidth{1.6}
  \Line(20,20)(30,20)
  \Line(20,-20)(30,-20)
  \Line(20,20)(20,-20)
\end{picture}}
& \hspace*{10mm} = &  \int \frac{{\mathfrak D}^dk_1 {\mathfrak
D}^dk_2}{P_0\left(k_1 + p_1\right) P_0\left( k_1
+p_1+p_2\right)P_0\left(k_2\right) P_0 \left(k_1-k_2\right)
P_m\left(k_1+p_3\right)} \, ,
\end{eqnarray}

\vspace*{5mm}
\noindent with

\begin{eqnarray}
\hbox{ \begin{picture}(0,0)(0,0)
\SetScale{1}
  \SetWidth{.5}
  \Line(-30,20)(30,20)
  \Line(-30,-20)(30,-20)
  \Line(-20,20)(-20,-20)
\CArc(0,42.5)(30,228,312)
  \SetWidth{1.6}
  \Line(20,20)(30,20)
  \Line(20,-20)(30,-20)
  \Line(20,20)(20,-20)
\end{picture}}
& \hspace*{16mm} = & \frac{1}{m^2} \sum_{i=-3}^0 A_i \, \vep + {\mathcal O}(\vep) \, ,
\end{eqnarray}

\begin{eqnarray} \nonumber
 A_{-3} &=& \frac{1}{32  (y+1)}
     \,, \\ \nonumber
 A_{-2} &=& \frac{1}{32  (y+1)} \; \Big(
  G(0;x)-2 G(1;x)-2 G(-1;y)+2
     \,\Big)\,, \\ \nonumber
 A_{-1} &=& \frac{1}{32  (y+1)} \;\Big[
 -3 \zeta(2)+G(0,0;x)-2 G(0,1;x)-2 G(1,0;x)+4 G(1,1;x) \\ \nonumber
  & & +G(0;x) (2-2 G(-1;y))-4
   G(-1;y)+G(1;x) (4 G(-1;y)-4) \\ \nonumber
  & & +4 G(-1,-1;y)-2 G(0,-1;y)+4
     \Big]\, , \\ \nonumber
 A_{0} &=& \frac{1}{32 (y+1)} \;\Big[
   -2 (3 \zeta(2)+4 \zeta(3)-4)+G(0,0,0;x)-2 G(0,0,1;x)-2 G(0,1,0;x)\\ \nonumber
  & & +4 G(0,1,1;x) -2 G\left(-1/y,0,0,x\right) +4 G\left(-1/y,0,1,x\right)+2 G\left(-1/y,1,0,x\right)\\ \nonumber
  & &-4 G\left(-1/y,1,1,x\right)  +2 G(-y,1,0;x) -4 G(-y,1,1;x)+G(1,1;x) (8-16 G(-1;y)) \\ \nonumber
  & &+G(0,0;x) (2-2 G(-1;y)) +2 (3 \zeta(2)-4) G(-1;y)-2 G\left(-1/y,0,x\right) G(-1;y) \\ \nonumber
  & & +4 G\left(-1/y,1,x\right) G(-1;y)-2 G(-y,0;x) G(-1;y)+4 G(-y,1;x)
   G(-1;y) \\ \nonumber
  & & +G(0,1;x) (4 G(-1;y)-4) +G(1,0;x) (8 G(-1;y)-4) \\ \nonumber
  & & +G(1;x) (2 (7 \zeta(2)-4)+8 G(-1;y))+8 G(-1,-1;y) \\ \nonumber
  & & +G(0;x) (-3 \zeta(2)-4 G(-1;y)+4 G(-1,-1;y) -2 G(0,-1;y)+4) \\ \nonumber
  & & -4G(0,-1;y)+G(-y;x) (2 G(0,-1;y)-4 G(-1,-1;y)) \\ \nonumber
  & & +\!G\left(-1/y,x\right) (-8 \zeta(2)-4 G(-1,-1;y)+2 G(0,-1;y)) \\
  & & -8 G(-1,-1,-1;y)\!+\!4 G(-1,0,-1;y)\!+\!4 G(0,-1,-1;y)-2 G(0,0,-1;y)
  \!\Big] \, .
\end{eqnarray}

We now consider the MIs involved in the calculation of the part of the amplitude
proportional to $N_h$. Topology Fig.~\ref{nhMIs}-(k) has two MIs:

\begin{eqnarray}
\hbox{\begin{picture}(0,0)(0,0)
\SetScale{1}
  \SetWidth{.5}
  \Line(-30,-20)(30,-20)
  \Line(-20,20)(20,-20)
  \Line(-30,20)(-20,20)
  \SetWidth{1.6}
  \Line(20,20)(30,20)
  \Line(20,-20)(30,-20)
  \Line(20,20)(20,-20)
\CArc(0,42.5)(30,228,312)
  \Line(-20,20)(30,20)
\end{picture}}
& \hspace*{10mm} = & \int \frac{{\mathfrak
D}^dk_1  {\mathfrak
D}^dk_2}{P_0\left(k_1 + p_1\right)
P_m\left(k_2\right)
P_m \left(k_1-k_2\right) P_m\left(k_1+p_3\right)} \, ,
\end{eqnarray}

\vspace*{5mm}
\noindent with

\begin{eqnarray}
\hbox{\begin{picture}(0,0)(0,0)
\SetScale{1}
  \SetWidth{.5}
  \Line(-30,-20)(30,-20)
  \Line(-20,20)(20,-20)
  \Line(-30,20)(-20,20)
  \SetWidth{1.6}
  \Line(20,20)(30,20)
  \Line(20,-20)(30,-20)
  \Line(20,20)(20,-20)
\CArc(0,42.5)(30,228,312)
  \Line(-20,20)(30,20)
\end{picture}}
& \hspace*{16mm} = & \sum_{i=-2}^1 A_i \, \vep + {\mathcal O}\left(\vep\right)
\, ,
\end{eqnarray}
\begin{eqnarray} \nonumber
 A_{-2} &=& \frac{1}{32}, \\ \nonumber
 A_{-1} &=& \frac{1}{32 y} \; \left[ 5 y-2 (y+1) G(-1;y) \right], \\ \nonumber
 A_{0} &=& \frac{1}{32 y (y+1)} \; \left[
 -10 G(-1;y) (y+1)^2+4 G(-1,-1;y) (y+1)^2-4 G(0,-1;y) (y+1) \right. \\ \nonumber
  && \left. -4 yG(0,0,-1;y)+y \left(19 y-4 \zeta(3)+19\right)
  \right] \, , \\ \nonumber
 A_{1} &=& \frac{1}{32 y (y+1)} \; \Big[
 20 G(-1,-1;y) (y+1)^2-8 G(-1,-1,-1;y) (y+1)^2\\ \nonumber
  & & + 8 G(-1,0,-1;y) (y+1) + 8 G(0,-1,-1;y) (y+1)-8 (y-1) G(1,0,-1;y) (y+1)\\ \nonumber
  & &+4 (y-1) G(1;y) \zeta(2) (y+1)+4 (y-2) (2 y+1) G(0,0,-1;y)-8 y G(-1,0,0,-1;y)\\ \nonumber
  & &+8 y G(0,-1,0,-1;y) +8 y G(0,0,-1,-1;y) -12 y G(0,0,0,-1;y)\\ \nonumber
  & &  +16 y G(0,1,0,-1;y)-8 y G(0,1;y) \zeta(2)+4 G(0,-1;y) \left(2 \zeta(2) y-5 y-5\right)\\ \nonumber
  & &-2 G(-1;y) \left(2 \zeta(2) y^2+19 y^2+4 \zeta(3) y+38 y-2 \zeta(2)+19\right) +\frac{y}{5}  \left(6 \zeta(2)^2+325 y \right.\\ 
  & &\left.+20 y \zeta(3)-40 \zeta(3)+325\right) \Big] \, .
\end{eqnarray}

As second MI for the topology Fig.~\ref{nhMIs}-(k), we chose

\begin{eqnarray}
\hbox{\begin{picture}(0,0)(0,0)
\SetScale{1}
  \SetWidth{.5}
  \Line(-30,-20)(30,-20)
  \Line(-20,20)(20,-20)
  \Line(-30,20)(-20,20)
  \SetWidth{1.6}
  \Line(20,20)(30,20)
  \Line(20,-20)(30,-20)
  \Line(20,20)(20,-20)
\CArc(0,42.5)(30,228,312)
  \Line(-20,20)(30,20)
  \GCirc(0,0){2.5}{0}
\end{picture}}
& \hspace*{10mm} = & \int \frac{{\mathfrak
D}^dk_1  {\mathfrak
D}^dk_2}{P_0^2\left(k_1 + p_1\right)
P_m\left(k_2\right)
P_m \left(k_1-k_2\right) P_m\left(k_1+p_3\right)} \, ,
\end{eqnarray}

\vspace*{5mm}
\noindent where

\begin{eqnarray}
\hbox{\begin{picture}(0,0)(0,0)
\SetScale{1}
  \SetWidth{.5}
  \Line(-30,-20)(30,-20)
  \Line(-20,20)(20,-20)
  \Line(-30,20)(-20,20)
  \SetWidth{1.6}
  \Line(20,20)(30,20)
  \Line(20,-20)(30,-20)
  \Line(20,20)(20,-20)
\CArc(0,42.5)(30,228,312)
  \Line(-20,20)(30,20)
  \GCirc(0,0){2.5}{0}
\end{picture}}
& \hspace*{16mm} = & \frac{1}{m^2} \sum_{i=-2}^1 A_i \, \vep^i+
{\mathcal O}\left(\vep^2\right) \, ,
\end{eqnarray}

\begin{eqnarray} \nonumber
 A_{-2} &=& -\frac{1}{16  (y+1)}, \\ \nonumber
 A_{-1} &=& \frac{y-1}{16  y (y+1)}  G(-1;y) \, , \\ \nonumber
 A_{0} &=& \frac{y-1}{8  y (y+1)^2} \;\Big[
 (y+1) G(-1;y) - (y+1) G(-1,-1;y) \\ \nonumber
  & & - G(0,-1;y)+\frac{2 y}{1-y} \left(y-\zeta(2)+1\right)
  \,\Big] \, , \\ \nonumber
A_{1}  &=&  \frac{y-1}{8  y (y+1)^2} \; \Big[
  -2 (y+1) G(-1,-1;y)+2 G(0,-1;y)+2 (y+1) G(-1,-1,-1;y)\\ \nonumber
  & & -2 G(-1,0,-1;y)-2 G(0,-1,-1;y)-2 (y-1) G(0,0,-1;y)\\ \nonumber
  & & +2 (y-1) G(1,0,-1;y)-(y-1) G(1;y) \zeta(2)+G(-1;y) \left(\zeta(2) y+2 y-\zeta(2)+2\right)\\ 
  & &-\frac{y}{y-1} \left(12 \ln (2) \zeta(2)-4 \zeta(2)+y \zeta(3)-8 \zeta(3)\right)
  \Big] \, .
\end{eqnarray}

The box  topology Fig.~\ref{nhMIs}-(l), has also two MIs: the first one is

\begin{eqnarray}
\hbox{\begin{picture}(0,0)(0,0)
\SetScale{1}
  \SetWidth{.5}
  \Line(-30,20)(30,20)
  \Line(-30,-20)(30,-20)
  \Line(-20,20)(-20,-20)
  \SetWidth{1.6}
  \Line(20,20)(30,20)
  \Line(-20,20)(20,20)
  \Line(20,-20)(30,-20)
  \Line(20,20)(20,-20)
\CArc(0,42.5)(30,228,312)
\end{picture}}
& \hspace*{10mm} = & \int \frac{{\mathfrak
D}^dk_1  {\mathfrak
D}^dk_2}{P_0\left(k_1 + p_1\right)
P_0\left( k_1 +p_1+p_2\right)
P_m\left(k_2\right)
P_m \left(k_1-k_2\right) P_m\left(k_1+p_3\right)} \, ,
\end{eqnarray}

\vspace*{5mm}
\noindent with

\begin{eqnarray}
\hbox{\begin{picture}(0,0)(0,0)
\SetScale{1}
  \SetWidth{.5}
  \Line(-30,20)(30,20)
  \Line(-30,-20)(30,-20)
  \Line(-20,20)(-20,-20)
  \SetWidth{1.6}
  \Line(20,20)(30,20)
  \Line(-20,20)(20,20)
  \Line(20,-20)(30,-20)
  \Line(20,20)(20,-20)
\CArc(0,42.5)(30,228,312)
\end{picture}}
& \hspace*{16mm} = & \frac{1}{m^2} \sum_{i=-3}^0 A_i \, \vep^i+
{\mathcal O}\left(\vep\right) \, ,
\end{eqnarray}

\begin{eqnarray} \nonumber
 A_{-3} &=& \frac{1}{32  (y+1)} \, , \\ \nonumber
 A_{-2} &=& \frac{1}{32  (x-1) (y+1)} \; \Big[
  -2 G(-1;y) (x-1)+2 (x-1)-(1+x) G(0;x)
     \Big] \, , \\ \nonumber
 A_{-1} &=& \frac{1}{32  (x-1) (y+1)} \Big[  3 \zeta(2) x+4 x+6 (x+1)
 G(-1,0;x)+(-5 x-1) G(0,0;x)\\ \nonumber
  & &-4 (x-1) G(-1;y) +G(0;x) (2 (x+1) G(-1;y)-2 (x+1))+
  4 (x-1) G(-1,-1;y) \\ \nonumber
  & & -2 (x-1) G(0,-1;y)+3 \zeta(2)-4
     \Big] \, , \\ \nonumber
 A_{0} &=& \frac{1}{32  (x-1) (y+1)} \Big[  -36 (x+1) G(-1,-1,0;x)+18 (x+1)
  G(-1,0,0;x)\\ \nonumber
  & &+6 (5 x+1) G(0,-1,0;x)
   +(-5 x-1) G(0,0,0;x)-2 (5 x-3) G(1,0,0;x) \\ \nonumber
  & & +4 (x+1) G(1,1,0;x)+2 (x+1) G\left(-1/y,0,0,x\right)-2 (x+1) G\left(-1/y,1,0,x\right)\\ \nonumber
  & &-\!2 (x\!+\!1) G(-y,1,0;x) -4 (x\!+\!1) G(1,0;x) G(-1;y)\!+\!2 (x\!+\!1) G\left(-1/y,0,x\right) G(-1;y) \\ \nonumber
  & & +2 (x+1) G(-y,0;x) G(-1;y)+G(-1,0;x) (12 (x+1)-12 (x+1) G(-1;y)) \\ \nonumber
  & & +G(0,0;x) (2 (5 x+1) G(-1;y)-2 (5 x+1))+8 (x-1) G(-1,-1;y) \\  \nonumber
  & & -4 (x-1) G(0,-1;y)-2 (x+1) G(-y;x) G(0,-1;y)-8 (x-1) G(-1,-1,-1;y) \\ \nonumber
  & & +4 (x-1) G(-1,0,-1;y)+4 (x-1) G(0,-1,-1;y)-2 (3 x-1) G(0,0,-1;y) \\  \nonumber
  & & -18 (x\!+\!1) G(-1;x) \zeta(2)-4 (x\!+\!1) G(1;x) \zeta(2)-2 G(-1;y) \left(3 \zeta(2) x\!+\!4 x\!+\!3 \zeta(2)-4\right) \\ \nonumber
  & & +G(0;x) \big(15 \zeta(2) x-4 x+4 (x+1) G(-1;y)-4 (x+1) G(-1,-1;y) \\ \nonumber
    & & +2 (x\!+\!1) G(0,-1;y)+3 \zeta(2)-4 \big) +G\left(-1/y,x\right) \left(2 (x\!+\!1) G(0,-1;y)\!+\!4 (x\!+\!1) \zeta(2)\right) \\
  & & +2 \left(3 \zeta(2) x+12 \zeta(3) x+4 x+3 \zeta(2)+4 \zeta(3)-4\right)
  \,\Big] \, .
\end{eqnarray}

As second MI for topology Fig.~\ref{nhMIs}-(l) we chose

\begin{eqnarray}
\hbox{\begin{picture}(0,0)(0,0)
\SetScale{1}
  \SetWidth{.5}
  \Line(-30,20)(30,20)
  \Line(-30,-20)(30,-20)
  \Line(-20,20)(-20,-20)
  \SetWidth{1.6}
  \Line(20,20)(30,20)
  \Line(-20,20)(20,20)
  \Line(20,-20)(30,-20)
  \Line(20,20)(20,-20)
\CArc(0,42.5)(30,228,312)
  \GCirc(0,20){2.5}{0}
\end{picture}}
& \hspace*{10mm} = & \int \frac{{\mathfrak
D}^dk_1  {\mathfrak
D}^dk_2}{P_0\left(k_1 + p_1\right)
P_0\left( k_1 +p_1+p_2\right)
P_m^2\left(k_2\right)
P_m \left(k_1-k_2\right) P_m\left(k_1+p_3\right)} \, ,
\end{eqnarray}

\vspace*{5mm}
\noindent with

\begin{eqnarray}
\hbox{\begin{picture}(0,0)(0,0)
\SetScale{1}
  \SetWidth{.5}
  \Line(-30,20)(30,20)
  \Line(-30,-20)(30,-20)
  \Line(-20,20)(-20,-20)
  \SetWidth{1.6}
  \Line(20,20)(30,20)
  \Line(-20,20)(20,20)
  \Line(20,-20)(30,-20)
  \Line(20,20)(20,-20)
\CArc(0,42.5)(30,228,312)
  \GCirc(0,20){2.5}{0}
\end{picture}}
& \hspace*{16mm} = & \frac{1}{m^4} \sum_{i=-2}^0 A_i \, \vep^i+
{\mathcal O}\left(\vep\right) \, ,
\end{eqnarray}

\begin{eqnarray} \nonumber
 A_{-2} &=& \frac{x}{32  (x-1) (x+1) (y+1)} \; G(0;x) \, , \\ \nonumber
 A_{-1} &=& -\frac{x}{32 (x-1) (x+1) (y+1)} \Big[ 6 G(-1,0;x)-3 G(0,0;x)+2 G(0;x) G(-1;y)+3 \zeta(2)
     \Big]\, , \\ \nonumber
 A_{0} &=& -\frac{x}{32  (x-1) (x+1) (y+1)} \Big[18
  G(-1,0,0;x) -36 G(-1,-1,0;x)+18 G(0,-1,0;x)\\ \nonumber
  & &-\!3 G(0,0,0;x) \! -\!2 G(1,0,0;x)\!+\!4 G(1,1,0;x)\!+\!2 G\left(-1/y,0,0,x\right)\!-\!2 G\left(-1/y,1,0,x\right) \\ \nonumber
  & & -2 G(-y,1,0;x)-12 G(-1,0;x) G(-1;y)+6 G(0,0;x) G(-1;y) \\ \nonumber
  & & -4 G(1,0;x) G(-1;y)+2 G\left(-1/y,0,x\right) G(-1;y)+2 G(-y,0;x) G(-1;y) \\ \nonumber
  & & -2 G(-y;x) G(0,-1;y)-2 G(0,0,-1;y)-18 G(-1;x) \zeta(2)-4 G(1;x) \zeta(2) \\ \nonumber
  & & -6 G(-1;y) \zeta(2)+G\left(-1/y,x\right) \left(2 G(0,-1;y)+4 \zeta(2)\right) \\
  & & +G(0;x) \left(-4 G(-1,-1;y)+2 G(0,-1;y)+9 \zeta(2) \right)+16 \zeta(3)
     \,\Big] \, .
\end{eqnarray}


\end{document}